\documentclass[english,11pt,a4paper,DIV=16,numbers=noenddot]{scrartcl}
\pdfoutput=1
\usepackage[utf8]{inputenc}
\usepackage{epsf}
\usepackage{amsmath, amssymb, amsthm}
\usepackage{amsfonts} 
\usepackage{graphicx}
\usepackage{listings}
\usepackage{array}
\usepackage{dcolumn}
\usepackage{mathtools}
\usepackage{scalefnt,ulem,soul}
\usepackage{cite}
\usepackage{color}
\usepackage{enumitem}
\usepackage{float}
\usepackage[table]{xcolor}

\usepackage{etoolbox}

\makeatletter
\pretocmd\start@align
{%
  \let\everycr\CT@everycr
  \CT@start
}{}{}
\apptocmd{\endalign}{\CT@end}{}{}
\makeatother

\let\theparentequation\theequation
\patchcmd{\theparentequation}{equation}{parentequation}{}{}

\renewenvironment{subequations}[1][]{
  \refstepcounter{equation}%
  \setcounter{parentequation}{\value{equation}}
  \setcounter{equation}{0}
  \def\theequation{\theparentequation\alph{equation}}%
  \let\parentlabel\label
  \ifx\\#1\\\relax\else\label{#1}\fi
  \ignorespaces
}{%
  \setcounter{equation}{\value{parentequation}}
  \ignorespacesafterend
}

\newcommand*{\nextParentEquation}[1][]{
  \refstepcounter{parentequation}
  \setcounter{equation}{0}
  \ifx\\#1\\\relax\else\parentlabel{#1}\fi
}

\makeatletter
\newlength{\fnhskip}
\renewcommand\@makefntext[1]{
  \settowidth{\fnhskip}{\@makefnmark}
  \leftskip=\fnhskip
  \hskip-\fnhskip
  \@makefnmark#1
}
\makeatother
\usepackage{setspace}

\usepackage[numbers,sort&compress]{natbib}
\def\NAT@spacechar{c}

\usepackage[
colorlinks=true
,urlcolor=blue
,anchorcolor=blue
,citecolor=blue
,filecolor=blue
,linkcolor=blue
,menucolor=blue
,pagecolor=blue
,linktoc=all
,unicode
,backref=page
]{hyperref}
\usepackage{hypernat}

\usepackage[all]{hypcap}
\usepackage[format=plain, margin=0.5cm, font=footnotesize, labelfont=bf]{caption}

\usepackage{cleveref}

\usepackage{footnotebackref}

\addtokomafont{disposition}{\rmfamily}

\newrobustcmd*{\tocref}[1]{\hyperref[TOC]{\color{black}{#1}}}
\renewcommand*{\backref}[1]{}
\renewcommand*{\backrefalt}[4]{%
  \ifcase #1%
  \or [p\,#2]%
  \else [pp\,#2]%
  \fi%
}

\newif\ifbackrefshowonlyfirst
\backrefshowonlyfirsttrue
\makeatletter
\let\BR@direct@old@hyper@natlinkstart\hyper@natlinkstart
\renewcommand*{\hyper@natlinkstart}{\phantomsection\BR@direct@old@hyper@natlinkstart}
\let\BR@direct@oldBR@citex\BR@citex
\renewcommand*{\BR@citex}{\phantomsection\BR@direct@oldBR@citex}%

\long\def\hyper@page@BR@direct@ref#1#2#3{\hyperlink{#3}{#1}}

\ifx\backrefxxx\hyper@page@backref
    \let\backrefxxx\hyper@page@BR@direct@ref
    \ifbackrefshowonlyfirst
    \fi
\else
    \ifbackrefshowonlyfirst
    \fi
\fi

\patchcmd{\Hy@backout}{Doc-Start}{\@currentHref}{}{\errmessage{I can't seem to patch backref}}
\makeatother

\apptocmd{\thebibliography}{\setlength{\itemsep}{0.02cm}}{}{}
\Crefname{figure}{Fig.}{Figs.}
\usepackage{placeins}

\let\theparentequation\theequation
\patchcmd{\theparentequation}{equation}{parentequation}{}{}

\BeforeTOCHead[toc]{{\pdfbookmark[1]{\contentsname}{toc}}}
\apptocmd{\thebibliography}{\scriptsize}{}{}
\let\OLDthebibliography\thebibliography
\renewcommand\thebibliography[1]{
  \OLDthebibliography{#1}
  \setlength{\parskip}{1pt}
  \setlength{\itemsep}{1pt plus 0.3ex}
}

\newcommand{\IE}{\textit{i.\,e.}}
\newcommand{\EG}{\textit{e.\,g.}}
\newcommand{\citere}[1]{Ref.\,\cite{#1}}
\newcommand{\citeres}[1]{Refs.\,\cite{#1}}

\newcommand{\code}[1]{\texttt{#1}}

\newcommand{\HB}{\code{HiggsBounds}}
\newcommand{\HS}{\code{HiggsSignals}}

\newcommand{\NMT}{\code{NMSSMTools}}
\newcommand{\CM}{\code{CheckMATE}}
\newcommand{\MG}{\code{MadGraph5}}
\newcommand{\abbrev}{\scalefont{.9}}
\newcommand{\eqn}[1]{Eq.\,(\ref{#1})}
\newcommand{\eqns}[1]{Eqs.\,(\ref{#1})}
\newcommand{\fig}[1]{Fig.\,\ref{#1}}
\newcommand{\figs}[1]{Figs.\,\ref{#1}}

\newcommand{\lhc}{{\abbrev LHC}}
\newcommand{\lep}{{\abbrev LEP}}

\newcommand{\sm}{{\abbrev SM}}
\newcommand{\vev}{{vev}}

\newcommand{\mssm}{{\abbrev MSSM}}
\newcommand{\nmssm}{{\abbrev NMSSM}}
\newcommand{\nnmmssm}{{\abbrev (N)MSSM}}
\newcommand{\munmssm}{{\abbrev $\mu$NMSSM}}
\newcommand{\gnmssm}{{\abbrev GNMSSM}}

\newcommand{\susy}{{\abbrev SUSY}}

\newcommand{\cp}{{\abbrev $\mathcal{CP}$}}

\newcommand{\MeV}{\textrm{MeV}}
\newcommand{\GeV}{\textrm{GeV}}
\newcommand{\TeV}{\textrm{TeV}}

\newcommand{\mue}{\mu_{\text{eff}}}
\newcommand{\mui}{\mu_{\text{inf}}}
\newcommand{\sign}{\ensuremath{\operatorname{sign}}}

\newcounter{notecount}
\makeatletter
\DeclareRobustCommand\em{%
  \@nomath\em \ifdim \fontdimen\@ne\font >\z@\scshape
  \else \slshape \fi}
\makeatother

\renewcommand{\emph}[1]{{\em #1}}

\hypersetup{
  pdfauthor={Wolfgang Gregor Hollik, Cheng Li, Gudrid Moortgat-Pick, Steven Paasch}
  ,pdftitle={Phenomenology of a Supersymmetric Model Inspired by Inflation}
  ,pdfsubject={}
  ,pdfkeywords={}
}

\title{
  \begin{flushright}
    {\textsf{\small
     DESY-20-059, TTP-2020-017, P3H-20-013}\\
    }
  \end{flushright}
	Phenomenology of a Supersymmetric Model Inspired by Inflation}

\author{Wolfgang Gregor Hollik${}^{a,b}$,
Cheng Li${}^{c}$,
Gudrid Moortgat-Pick${}^{d,c}$,
Steven Paasch${}^{c}$
  \\[1em]
\textit{${}^a$Institute for Nuclear Physics, Karlsruhe Institute of Technology,}\\
\textit{D-76021 Karlsruhe, Germany}\\[.7em]
\textit{${}^b$Institute for Theoretical Particle Physics, Karlsruhe Institute of Technology,}\\
\textit{D-76128 Karlsruhe, Germany}\\[.7em]
\textit{${}^c$DESY, Notkestra{\ss}e 85, D-22607 Hamburg, Germany}\\[1em]
\textit{${}^d$II. Institut f\"ur Theoretische Physik, Universit\"at Hamburg,}\\
\textit{Luruper Chaussee 149, D-22761 Hamburg, Germany}\\[1em]
{\small\texttt{wolfgang.hollik@kit.edu},
\texttt{cheng.li@desy.de},} \\[-0.3em]
\small{\texttt{gudrid.moortgat-pick@desy.de},
\texttt{steven.paasch@desy.de}}
}
\date{\today}

\addtokomafont{title}{\rmfamily}
\addtokomafont{section}{\rmfamily}
\addtokomafont{subsection}{\rmfamily}

\begin{document}
\maketitle
\thispagestyle{empty}

\begin{abstract}
The current challenges in High Energy Physics and Cosmology are to
build coherent particle physics models to describe the phenomenology
at colliders in the laboratory and the observations in the
universe. From these observations, the existence of an inflationary
phase in the early universe gives guidance for particle physics
models. We study a supersymmetric model which incorporates
successfully inflation by a non-minimal coupling to supergravity and
shows a unique collider phenomenology. Motivated by experimental data,
we set a special emphasis on a new singlet-like state at \(97\,\GeV\)
and single out possible observables for a future linear collider that
permit a distinction of the model from a similar scenario without
inflation. We define a benchmark scenario that is in agreement with
current collider and Dark Matter constraints, and study the influence
of the non-minimal coupling on the phenomenology.  Measuring the
singlet-like state with high precision on the percent level seems to
be promising for resolving the models, even though the Standard
Model-like Higgs couplings deviate only marginally.  However, a
hypothetical singlet-like state with couplings of about \(20\,\%\)
compared to a Standard Model Higgs at \(97\,\GeV\) encourages further
studies of such footprint scenarios of inflation.
\end{abstract}

\newpage
\tableofcontents
\begin{center}
\hrule
\end{center}
\section{Introduction} \label{sec:intro}
Supersymmetry (\susy{}) remains a valid conceptual extension beyond
the Standard Model (\sm{}) of particle physics, although there have
not yet been any direct signs of superpartners detected in
proton--proton collisions even at 13 TeV center of mass energy at the
run 2 of the Large Hadron Collider (LHC).  Nevertheless, light SUSY
states from the electroweak sector cannot be excluded, not even at
relatively low masses. The concept of \susy{} as a space-time symmetry
is mathematically sound, phenomenologically beautiful and connects the
fundamental forces of the SM with gravity. In supergravity, moreover,
a non-minimal gravitational coupling of the Higgs particle content
leads to a successful embedding of inflation in the early
universe~\cite{Einhorn:2009bh, Ferrara:2010yw, Ferrara:2010in,
  Lee:2010hj}. The basic Higgs phenomenology of this variation of a
Next-to-Minimal Supersymmetric Standard Model (NMSSM) has been
described in some detail in~\cite{Hollik2019}, where it has been
argued, that the main effect of the non-minimal supergravity coupling
might be visible in a precise study of a singlet-like Higgs state that
has to be discovered at the LHC or future lepton colliders.

Especially the option of a light additional Higgs state at
\(97\,\GeV\) as favoured by some observational hints at the Large
Electron--Positron Collider (LEP)~\cite{Abbiendi:2002qp,
  Barate:2003sz, Schael:2006cr} and the LHC~\cite{Sirunyan:2018aui,
  Sirunyan:2018zut, ATLAS:2018xad}, which can be present in many
singlet extended models~\cite{Dermisek:2007ah, Belanger:2012tt,
  Cao:2016uwt, Domingo:2018uim, Biekotter:2017xmf, Biekotter:2019kde,
  Biekotter:2019gtq, Cao:2019ofo, Cline:2019okt, Choi:2019yrv,
  Richard:2020jfd, AguilarSaavedra:2020wmg, Biekotter:2020cjs}, is an
intriguing case study also for the inflation-inspired model. We want
to state that the existence or nonexistence of such a light Higgs is
neither unique to the model which is going to be studied in the
current work, nor is it a special feature of it.\footnote{Especially
  the existence of a singlet-like Higgs state below \(125\,\GeV\) can
  be present in certain parameter regions of the \nmssm,
  see~\citere{Ellwanger:2015uaz}.}  However, it is interesting to
connect to new light bosons as they could be studied with
unprecedented precision in future $e^+e^-$-colliders, for instance at
the International Linear Collider (ILC) with an initial low center of
mass energy of \(250\,\GeV\). Thus, we are going to put special
emphasis on the $e^+e^-$-collider phenomenology of a benchmark point
which comprises such a scalar boson below \(100\,\GeV\).

This paper is structured as follows: first, we briefly review the
supersymmetric model motivated by inflation in
Sec.~\ref{sec:framework} which has been introduced in
\citeres{Ferrara:2010yw, Ferrara:2010in}. The phenomenology of the the
Higgs and of the electroweakino sector has already been discussed in
detail in \citere{Hollik2019} to which we closely relate here. Second,
we perfom a scan of the relevant model parameters from which we
extract a benchmark scenario which is discussed in more detail in
Sec.~\ref{sec:discussion} and discuss the phenomenology of such a
scenario. Finally, our conclusions are presented in
Sec.~\ref{sec:concl}.

\section{Theoretical Framework} \label{sec:framework}
The model with successful early universe inflation in the context of
superconformal supergravity~\cite{Ferrara:2010yw, Ferrara:2010in,
  Lee:2010hj} can be embedded in the general \nmssm{} (\gnmssm{}) as
reviewed in \citere{Ellwanger2010}.  In order to drive inflation, a
non-minimal coupling of a Higgs bilinear to gravity is needed, which
has been shown to be the gauge invariant product \(\hat H_u \cdot \hat
H_d\), as pointed out in \citere{Einhorn:2009bh}. The singlet
superfield is needed to stabilise the inflationary
direction~\cite{Ferrara:2010yw, Ferrara:2010in, Lee:2010hj}. At low
(electroweak) energies, the superpotential is given by the
superpotential of the \(\mathbb{Z}_3\)-invariant \nmssm{} plus an
additional \(\mu\)-term like in the Minimal Supersymmetric Standard
Model (\mssm) \(\mu\,\hat H_u \cdot \hat H_d\). This parameter we name
for clarity \(\mui\) and the model thus ``\(\mu\)-extended'' \nmssm{}
or short \munmssm.  In contrast to the \(\mathbb{Z}_3\)-invariant
\nmssm{}, there is no accidental \(\mathbb{Z}_3\) symmetry prohibiting
certain terms in the superpotential of the \gnmssm{} like the
\(\mu\)-term for the two Higgs doublet superfields and the mass and
tadpole term for the singlet superfield.  The \(\mu\)-term breaks the
\(\mathbb{Z}_3\) symmetry of the \nmssm{} and thus also
non-\(\mathbb{Z}_3\)-invariant terms in the soft \susy{} breaking
sector are supposed to be present.  Nevertheless, due to breaking of
the superconformal symmetry by only the gravitational coupling, the
superpotential does not introduce the mass and tadpole term for the
singlet. The soft breaking terms can always be redefined in a way that
only the couplings introduced below are relevant.

The superpotential of the \munmssm{} is given by
\begin{align} \label{eq:superpot}
\mathcal{W}_{\mu\text{NMSSM}}=&(\lambda\hat{S}+\mui)\hat{H}_u\cdot\hat{H}_d+\frac{\kappa}{3}\hat{S}^3
+ \mathcal{W}_\text{Yukawa}\,,
\end{align}
where the extra \(\mu\)-term is related to the non-minimal
supergravity coupling \(\chi\) via the gravitino mass as \(\mui =
\frac{3}{2}m_{3/2}\chi\). The Yukawa terms are the same as in the
\nnmmssm. Chiral superfields are denoted with a hat, where \(\hat H_u\)
and \(\hat H_d\) are the up- and down-type Higgs doublet,
respectively, and \(\hat S\) the singlet superfield. The corresponding
soft \susy{} breaking Lagrangian is given by
\begin{align}\label{eq:break}
  \begin{split}
  -\mathcal{L}_{\text{soft}} &= \left[A_\lambda\,\lambda\,S\,H_u\cdot
    H_d + \frac{1}{3}\,A_\kappa\,\kappa\,S^3 + B_\mu\,\mu\,H_u\cdot
    H_d + \text{h.\,c.}\right]\\ &\quad + m_{H_d}^2\,\lvert
  H_d\rvert^2 + m_{H_u}^2\,\lvert H_u\rvert^2 + m_s^2\,\lvert
  S\rvert^2\,.
\end{split}
\end{align}
The soft \susy{} breaking Higgs masses can be related to the
electroweak symmetry breaking conditions and are no free
parameters. The \(B_\mu\) terms play a subdominant role and can be set
to zero throughout this work.

After electroweak symmetry breaking, the scalar components of the
three Higgs superfields acquire vacuum expectation values (\vev{}s)
$v_u$, $v_d$ and $v_s$. We expand these fields around the vacuum
configuration and write:
\begin{gather} \label{eq:Higgsfields}
H_u=\begin{pmatrix}
h_u^+\\
h_u
\end{pmatrix}=
\begin{pmatrix}
\eta_u^+\\
v_u+\frac{1}{\sqrt{2}}(\sigma_u+i\phi_u)
\end{pmatrix},\quad H_d=\begin{pmatrix}
h_d\\
h_d^-
\end{pmatrix}=
\begin{pmatrix}
v_d+\frac{1}{\sqrt{2}}(\sigma_d+i\phi_d)\\
\eta_d^-
\end{pmatrix}\notag\\
S=v_s+\frac{1}{\sqrt{2}}(\sigma_s+i\phi_s).
\label{eq:hf}
\end{gather}
The ratio of the two doublet \vev{}s defines the parameter \(\tan\beta
= v_u/v_d\), where \(v = \sqrt{v_u^2 + v_d^2} = 174\,\GeV\)
corresponds to the \sm-\vev. Consequently, $v_u$ and $v_d$ are given
by \(v_u = v \sin\beta\) and \(v_d = v \cos\beta\).  The \vev{} of the
singlet field $S$ dynamically induces a $\mu$-term which we denote as
the \emph{effective} \(\mu\)-term, $\mue=\lambda v_s$. Although it
might be suggestive to combine the two \(\mu\)-terms as \(\mue \to
\mui+\mue\), they lead to different phenomenologies in the Higgs and
Neutralino sector, as has been pointed out in \citere{Hollik2019}.

Thus, we consider both $\mui$ and $\mue$ as independent free
parameters in our study. The consequent differences in the
phenomenology will be the crucial point of our discussion. The
Neutralino--Singlino mixing will also be affected by the interplay of
\(\mui\) and \(\mue\) and therewith the character of the contribution
to dark matter may vary. Since \(\mui\) is related to the gravitino
mass, dark matter might also be pure gravitino dark matter, see the
discussion in \citere{Hollik2019}.

According to the cosmological analysis \cite{Ferrara:2010in,
  Lee:2010hj}, the value of the non-minimal gravity coupling $\chi$
can be estimated to $\chi\simeq 10^5\lambda$. Thus, with \(\lambda >
0\), we also set $\mui$ to be non-negative.\footnote{One can always
  choose \(\lambda > 0\) and allow for negative \(\kappa\).}

\subsection{Higgs sector} \label{sec:Higgs}
The superpotential~\eqref{eq:superpot} and the soft-breaking
Lagrangian~\eqref{eq:break} together with the usual \(D\)-terms
(quartic Higgs couplings due to quadratic gauge couplings which do not
exist for the singlet) lead to the following scalar Higgs potential
(with \(B_\mu = 0\)):
\begin{align}
V_{\text{Higgs}}=\;&
\left(m_{H_d}^2+(\mu_{\text{inf}}+\lambda S)^2\right)|H_d|^2
+\left(m_{H_u}^2+(\mu_{\text{inf}}+\lambda S)^2\right)|H_u|^2 \nonumber \\
&\;+\left(\kappa S^2+\lambda H_u\cdot H_d\right)^2
+\frac{g_2^2}{2}|H_d^\dagger H_u|^2
+\frac{g_1^2+g_2^2}{8}\left(|H_d|^2-|H_u|^2\right)^2 \nonumber \\
&\;+m_S^2 S^2+2\lambda A_\lambda S H_u\cdot H_d+\frac{2}{3}\kappa A_\kappa S^3\,.
\label{eq:hv}
\end{align}
The mass terms finally arise from the second derivative with respect
to the component fields in \eqns{eq:Higgsfields} evaluated at the
vacuum. Note that the soft breaking terms \(m_{H_u}^2\), \(m_{H_d}^2\)
and \(m_S^2\) are fixed by the minimisation conditions for electroweak
symmetry breaking. For convenience, we list the mass matrix elements
of the scalar, pseudoscalar and charged Higgs matrices, \(M_S^2\),
\(M_P^2\), and \(M_C^2\), respectively, as worked out in
\citere{Hollik2019}; we only keep the contribution from \(\mui\) in
comparison with the \gnmssm{}:\footnote{We express in terms of the
  gauge boson masses
\[
  m_W^2=\frac{1}{2}g_2^2v^2,\quad
  m_Z^2=\frac{1}{2}(g_1^2+g_2^2) v^2\,.
\]}
\begin{subequations} \label{eq:mx1}
\begin{align}
&{M}_{S,11}^2=m_Z^2\cos^2\beta+\mue\left(\frac{\kappa}{\lambda}\mue+A_\lambda\right)\tan\beta \\
&{M}_{S,22}^2=m_Z^2\sin^2\beta+\mue\left(\frac{\kappa}{\lambda}\mue+A_\lambda\right)/\tan\beta \\
&{M}_{S,33}^2=\frac{\lambda^2v^2}{\mue}(\cos\beta\sin\beta A_\lambda-\mui)+\frac{\kappa}{\lambda}\mue\left(A_\kappa+4\frac{\kappa}{\lambda}\mu_{\text{eff}}\right) \\
&{M}_{S,12}^2={M}_{S,21}^2=(2v^2\lambda^2-m_Z^2)\cos\beta\sin\beta-\mue\left(\frac{\kappa}{\lambda}\mue+A_\lambda\right) \\
&{M}_{S,13}^2={M}_{S,31}^2=\lambda v\left(2(\mue+\mui)\cos\beta-\left(A_\lambda+2\frac{\kappa}{\lambda}\mue\right)\sin\beta\right) \\
&{M}_{S,23}^2={M}_{S,32}^2=\lambda v\left(2(\mue+\mui)\sin\beta-\left(A_\lambda+2\frac{\kappa}{\lambda}\mue\right)\cos\beta\right)\,,
\end{align}
\end{subequations}
\begin{subequations}
\begin{align}
&{M}_{P,11}^2=\mu_{\text{eff}}\left(\frac{\kappa}{\lambda}\mue+A_\lambda\right)\tan\beta \\
&{M}_{P,22}^2=\mu_{\text{eff}}\left(\frac{\kappa}{\lambda}\mue+A_\lambda\right)/\tan\beta \\
&{M}_{P,33}^2=\frac{\lambda^2 v^2}{\mue}\Big((4\frac{\kappa}{\lambda}\mue+A_\lambda)\cos\beta\sin\beta-\mu_{\text{inf}}\Big)-3\frac{\kappa}{\lambda}\mue A_\kappa \\
&{M}_{P,12}^2={M}_{P,21}^2=\mu_{\text{eff}}\left(\frac{\kappa}{\lambda}\mue+A_\lambda\right) \\
&{M}_{P,13}^2={M}_{P,31}^2=-v\lambda\left(2\frac{\kappa}{\lambda}\mue-A_\lambda\right)\sin\beta \\
&{M}_{P,23}^2={M}_{P,32}^2=-v\lambda\left(2\frac{\kappa}{\lambda}\mue-A_\lambda\right)\cos\beta\,,
\label{eq:mx2}
\end{align}
\end{subequations}
\begin{subequations}
\begin{align}
&{M}_{C,11}^2=(m^2_W-v^2\lambda^2)\sin^2\beta+\mue\left(\frac{\kappa}{\lambda}\mu_{\text{eff}}+A_\lambda\right)\tan\beta \\
&{M}_{C,22}^2=(m^2_W-v^2\lambda^2)\cos^2\beta+\mue\left(\frac{\kappa}{\lambda}\mu_{\text{eff}}+A_\lambda\right)/\tan\beta \\
&{M}_{C,12}^2=(m^2_W-v^2\lambda^2)\sin\beta\cos\beta+\mue\left(\frac{\kappa}{\lambda}\mu_{\text{eff}}+A_\lambda\right)\,.
\label{eq:chms}
\end{align}
\end{subequations}
The pseudoscalar and charged mass matrix comprise one vanishing
eigenvalue each. These correspond to the would-be-Goldstone
modes. Diagonalisation of those two matrices is easy and can be done
with a rotation by the angle \(\beta\). The charged Higgs mass is then
found to be given by the expression:
\begin{equation}
m_{H^\pm}^2=m_W^2-v^2\lambda^2+\frac{\mue}{\cos\beta\sin\beta}\left(\frac{\kappa}{\lambda}\mue+A_\lambda\right)\,,
\label{eq:ma}
\end{equation}
from which we can resolve for \(A_\lambda\) and use \(m_{H^\pm}\) as
input parameter to replace the appearance of \(A_\lambda\) in the
model. We then can use the relation
\begin{equation}
\mue\left(\frac{\kappa}{\lambda}\mue+A_\lambda\right)=(m_{H^\pm}^2-m_W^2+v^2\lambda^2)\cos\beta\sin\beta
\label{eq:al}
\end{equation}
to cancel out the $\kappa$ and $\mue$ dependences in \eqns{eq:mx2} and
\eqref{eq:chms}. Furthermore, if we fix $m_{H^\pm}$ to a large value
\(m_{H^\pm}^2 \gg v^2\), the heaviest neutral Higgs bosons both for
CP-even and CP-odd case are basically independent of $\kappa$, $\mue$
and $\mui$; \IE{} the heavy mass eigenvalues are dominantly controlled
by \(m_{H^\pm}\).  We are in general left with the following free
parameters in our study:
\begin{equation}
\tan\beta, \quad \lambda,\quad \kappa,\quad \mue,\quad
\mui,\quad A_\kappa,\quad m_{H^\pm}.
\end{equation}
In the following, we treat both $\tan\beta$ and \(m_{H^\pm}\) as fixed
input parameters that are kept to some experimentally allowed
value. By this choice, the matrix elements $M_{S,P,11}^2$,
$M_{S,P,22}^2$ and $M_{S,P,12}^2$ do not vary under variation of the
other inputs. We are interested in the effect of the inflation
specific parameters, for which \(\tan\beta\) and \(m_{H^\pm}\) play a
subleading role and have rather the same influence as in the usual
\nmssm. The further elements $M_{S,P,13}^2$, $M_{S,P,23}^2$ and
$M_{S,P,33}^2$ are then mainly controlled by the parameter
combinations $\frac{\kappa}{\lambda}\mu_{\text{eff}}$ and the sum
$\mu_{\text{eff}}+\mu_{\text{inf}}$ aside from \(m_{H^\pm}\).  Thus,
the properties of the light neutral Higgs states at tree level are
dominated by these two combinations, although the other free
parameters $\lambda$, $A_\kappa$, and $\mu_{\text{inf}}$ can influence
the mass matrices.

From the diagonalisation, we retrieve the Higgs mixing parameters
$S_{ij}$, $P_{ij}$ and $C_{ij}$ for the scalar, pseudoscalar and
charged cases, respectively. The diagonal matrices are found as
\(\boldsymbol{\tilde M}_S^2 = \boldsymbol{S}^\dag \boldsymbol{M}_S^2
\boldsymbol{S}\), \(\boldsymbol{\tilde M}_P^2 = \boldsymbol{P}^\dag
\boldsymbol{M}_P^2 \boldsymbol{P}\), and \(\boldsymbol{\tilde M}_C^2 =
\boldsymbol{C}^\dag \boldsymbol{M}_C^2 \boldsymbol{C}\).  With the
mixing matrices, the Higgs couplings to \sm{} particles can be
conveniently expressed and compared to the \sm{} values in terms of
``reduced'' couplings. So for example, reduced couplings of the
\(i\)-th scalar Higgs to bottom and top quarks are given by:
\begin{equation} \label{eq:topbotcoup}
\frac{g_{h_{i}b\bar{b}}}{g_{H_{\text{SM}}b\bar{b}}}=\frac{S_{i1}}{\cos\beta}\,,
\qquad\qquad
\frac{g_{h_{i}t\bar{t}}}{g_{H_{\text{SM}}t\bar{t}}}=-\frac{S_{i2}}{\sin\beta}\,,
\end{equation}
and the reduced coupling to gauge bosons reads:
\begin{equation}
\frac{g_{h_{i}ZZ}}{g_{H_{\text{SM}}ZZ}}=\frac{g_{h_{i}W^+W^-}}{g_{H_{\text{SM}}W^+W^-}}=\cos\beta S_{i1}+\sin\beta S_{i2}\,.
\label{eq:cp}
\end{equation}
Note, that in the \munmssm, as well as the \nmssm, the reduced gauge
boson couplings for \(Z\) and \(W\) are the same at the tree level. In
the course of this work, we explicitly focus on the Higgsstrahlung
process at lepton colliders, for which the cross section is controlled
by the Higgs coupling to vector bosons $g_{HVV}$.

Although the reduced couplings from above\footnote{The reduced
  couplings are defined at tree level. Radiative corrections are
  implemented in the mixing matrix elements \(S_{ij}\) as they are
  defined from the loop-corrected mass matrices in \NMT.} cannot be
directly probed by experiment, they give important information for the
production and decay cross sections. In the so-called
\(\kappa\)-framework, effective Higgs couplings are determined from
measured rates in the relevant channels. The reduced couplings are
then found from ratios of cross section times branching ratios. The
coupling-strength modifiers \(\kappa\) are not to be identified with
the reduced couplings. However, under certain assumptions like a small
width the difference is negligible for a leading order analysis. In
case the production and decay can be factorised, the coupling
modifiers factor out as
\begin{equation}
  \sigma(X \to H) \operatorname{Br}(H \to f) = \kappa_X^2 \;
  \kappa_f^2 \; \sigma^\text{SM}_X \;
  \frac{\Gamma_f^\text{SM}}{\Gamma_H(\kappa_X^2, \kappa_f^2)}\,,
\end{equation}
with the \sm{} production cross section \(\sigma_X^\text{SM}\) and the
partial decay width for the \sm{} Higgs \(\Gamma_f^\text{SM}\) into a
certain final state \(f\). \(\Gamma_H (\kappa_X^2, \kappa_f^2)\) is
the total width in presence of the coupling modifiers \(\kappa_X\) and
\(\kappa_f\). The individual modified coupling strengths can be found
as the ratios
\begin{equation}
  \kappa_X^2 = \frac{\sigma_X}{\sigma_X^\text{SM}}
  \quad \text{and} \quad
  \kappa_f^2 = \frac{\Gamma_f}{\Gamma_f^\text{SM}}\,.
\end{equation}
Note that in general higher order accuracy is lost and the \(\kappa\)
can be more complicated functions of the reduced couplings. The latter
is especially important for the modified couplings to gluons and
photons~\cite{Heinemeyer:2013tqa}. This has to be included in a
correct study of the modified couplings.

For our numerical studies, we refer to the \texttt{NMSSMTools} package
\cite{Ellwanger:2004xm, Ellwanger:2006rn, Das:2011dg, Domingo:2015qaa}
as spectrum generator and for calculations of some crucial
observables\footnote{We are using the highest possible precision
  implemented in \NMT{} for the \gnmssm: full one loop top/bottom
  contribution plus leading logarithmic two loop top/bottom and leading
  logarithmic one loop electroweak corrections~\cite{NMSSMTools-www}
  to the neutral Higgs masses and mixings.} that are given below.
Although \NMT{} does not provide the input for the \munmssm, but rather
the \gnmssm, we can redefine the input parameters in a way that is
compatible with the \munmssm. Note, that in the \gnmssm, out of the
three \(\mathbb{Z}_3\)-breaking parameters in the superpotential, one
can always be eliminated by redefinition of the others. Since in
\texttt{NMSSMTools} the input list does not contain the general \(\mu\)
parameter which corresponds to \(\mui\), we have transferred the effect
to the other parameters and redefine the overall inputs by the following
replacement list:
\begin{subequations}
\begin{align}
        \mue&\to\mue+\mui \,, \\
        \kappa&\to\kappa\frac{\mue}{\mue+\mui} \,, \\
        \mu'&\to 0\,, \\
        \xi_F&\to 0\,, \\
        \xi_S&\to\frac{\lambda}{\mue}(v^2\mui(\mue+\mui)-v_u v_d  A_\lambda\mui)\,, \\
        m^2_3&\to -\mui(A_\lambda+\frac{\kappa}{\lambda}\mue)\,, \\
        m_S'^2&\to -2\frac{\kappa\lambda \mui}{\mue+\mui}v_uv_d\,.
\end{align}
\end{subequations}
By this redefinitions, also the additional soft-breaking terms are
involved and thus all effects and arising singularities in the quantum
corrections are appropriately taken care of. The superpotential
parameters \(\mu'\) and \(\xi_F\), cf.~\citere{Ellwanger2010}, are
protected by supersymmetry and can be set to zero at all
scales.\footnote{The notation of the \gnmssm{} parameters in
  \citere{Hollik2019} is \(\mu' = \nu\), \(\xi_F = \xi\), \(\xi_S =
  \xi C_\xi\), \(m_3^2 = \mu B_\mu\), \({m'}_S^2 = \nu B_\nu\).}

\subsection{Gaugino and chargino sector}
In the \munmssm{}, the Higgsino mass parameter is given by
\((\mue+\mui)\) instead of \(\mue\) in the \nmssm. In contrast to the
\nmssm, however, the singlino mass is driven by a different
combination. The symmetric mass matrices for neutralinos and charginos
are given by (see \EG{} \citere{Hollik2019} and references therein)
\begin{align}
\label{eq:neutralinomassmatrix}
M_{{\tilde\chi}^0} &=
\begin{pmatrix}
M_{1} & 0     & -m_{Z}\sin\theta_{\text{w}} \cos\beta & m_{Z}\sin\theta_{\text{w}}\sin\beta     & 0 \\
\cdot & M_{2} & m_{Z}\cos\theta_{\text{w}}\cos\beta  & -m_{Z}\cos\theta_{\text{w}}\sin\beta   & 0 \\
\cdot & \cdot & 0                                                       & -(\mui+\mue)          & -\lambda \upsilon \sin\beta \\
\cdot & \cdot & \cdot                                           & 0                                                     & -\lambda \upsilon \cos\beta \\
\cdot & \cdot & \cdot                                           & \cdot                                                 & 2 \frac{\kappa}{\lambda} \mue \\
\end{pmatrix}
\,,\\[.2em]
M_{{\tilde\chi}^{\pm}} &=
\begin{pmatrix}
M_{2}                                                   & \sqrt{2}m_{W} \sin\beta    \\
\sqrt{2}m_{{W}} \cos\beta  & \mui+\mue\\
\end{pmatrix}\,,
\end{align}
where $\theta_{\text{w}}$ is the weak mixing angle and \(M_{1,2}\) the
soft \susy{} breaking gaugino masses for the bino and wino,
respectively. The matrices are given in the basis of gauge eigenstates, where:
\begin{equation}
  ({\tilde\psi}^{0})^{T}=(\tilde{B}^{0}, \tilde{W}_{3}^{0},
  \tilde{h}_{d}^{0}, \tilde{h}_{u}^{0},\tilde{s}^{0})\,, \quad
  ({\tilde\psi}^{+})^{T}=( \tilde{W}^{+}, \tilde{h}_{u}^{+}) \quad
  \text{and} \quad
  ({\tilde\psi}^{-})^{T}=( \tilde{W}^{-}, \tilde{h}_{d}^{-})\,,
\end{equation}
with the bino $\tilde{B}^{0}$, the neutral and charged wino components
$\tilde{W_{3}}^{0}$ and $\tilde{W}^{\pm}$, the charged and neutral
higgsino components $\tilde{h}_{u,d}^{\pm}$ and $\tilde{h}_{u,d}^{0}$,
and the singlino component $\tilde{s}^{0}$. The mass eigenstates are
denoted by the neutralinos \({\tilde\chi}^0_{1-5}\) and charginos
\({\tilde\chi}^\pm_{1,2}\).

One can see that the mass of the higgsino component is driven by the
sum $\mui+\mui$, while the mass scale of the singlino component is
driven by $ \frac{\kappa}{\lambda} \mue$. Since the singlino mass is
the only matrix element that containts the parameter \(\kappa\) at the
tree level, one may use this to reweight any relative shift between
\(\mue\) and \(\mui\) by a change of \(\kappa\) in order to keep the
neutralino spectrum under variation of \(\mui\). This rescaling
procedure has been described in \citere{Hollik2019} and will be also
used in the following to tackle the effect of \(\mui\) in the model.

\section{Phenomenological discussion} \label{sec:discussion}
In this section we explore methods to experimentally distinguish the
\nmssm{} from the \munmssm. For this purpose, we perform a scan in the
\nmssm{} parameter space and select points passing a number of
experimental constraints. Based on one benchmark scenario we scan the
\munmssm{} parameter space for points with a similar mass spectrum
within an interval of a few \(\GeV\). We discuss experimental
observables like branching ratios and cross-sections to describe
features introduced by the parameter space of the \munmssm. Starting
from the \nmssm{} benchmark point, we show the effect from \(\mui\)
exclusively and the option to conceal the influence from this
parameter by a redefinition of others. Finally, we discuss methods to
experimentally distinguish both models.

\subsection{\nmssm{} benchmark points}
\label{sec:NMSSMpoints}
A full phenomenological discussion of the complete parameter space in
the \munmssm{} and \nmssm{} is a formidable task. We want to focus on
a certain feature in the Higgs mass spectrum comprising a light
neutral scalar boson. In order to achieve this, we have scanned for
points in the \nmssm{} parameter space having this feature and passing
the constraints given by \NMT{} version \code{5.5.2}
\cite{Ellwanger:2004xm, Ellwanger:2005dv, Ellwanger:2006rn} (\EG{}
certain collider observables and Dark Matter constraints), as well as
\HB{} version \code{5.3.2} \cite{Bechtle:2008jh, Bechtle:2013wla},
\HS{} version \code{2.5.0} \cite{Bechtle:2013xfa, Stal:2013hwa}, and
\CM{} version \code{2.0.26} \cite{deFavereau:2013fsa, Cacciari:2011ma,
  Cacciari:2005hq, Cacciari:2008gp, Read:2002hq, Drees:2013wra,
  Dercks:2016npn} for LHC analyses. As a consistency check, we also
interfaced \code{SModelS} version \code{1.2.4} \cite{Kraml:2013mwa,
  Khosa:2020zar} which has a complementary approach and uses
simplified models for direct collider bounds/searches. For the scan,
we have constrained ourselves to a variation of relevant parameters
only, where we keep less relevant \susy{} parameters at fixed
values.\footnote{Relevant for the study of \(\mui\) in the Higgs
  sector.} The codes are interfaced using the standard SUSY Les
Houches Accords (SLHA) according to \citeres{Skands:2003cj,
  Allanach:2008qq}. The values of all fixed parameters are given in
Tab.~\ref{tab:NMSSMfixedpar}. Besides the \sm{} parameters, we keep
the gaugino mass parameters $M_{1}$ and $M_{2}$ obeying the GUT
relation $M_{1} = \frac{5}{3} \frac{g^2_1}{g^2_2} M_{2}$ with $M_2 =
500\,\GeV$. 

\begin{table}[tb]
  \caption{Fixed \sm{} and \susy{} input parameters of the \nmssm{}
    scenario. The gaugino mass parameters are denoted as $M_{i}$ with
    $i = 1,2,3$ and the ratio of the electroweak vevs
    $\text{tan}\beta$. We have the trilinear soft-breaking sfermion
    term $A_{f_{3}}$ (the other \(A_{f_{1,2}}\) are set to zero), the
    sfermion mass $m_{\bar{f}_{L},\bar{f}_{R}}$ and also the pseudoscalar
    Higgs mass input $M_{A}$.}
\label{tab:NMSSMfixedpar}
\centering
\begin{tabular}{lll}
\hline
 &  &  \\
$m_{Z} = 91.187\,\GeV$ & $\alpha_{\text{em}}^{-1} = 127.92$ & $G_{F}=1.16637\cdot 10^{-5}\,\GeV^{-2}$ \\
 &  & \\
$M_{1} = 239\,\GeV$ & $M_{2} = 500\,\GeV$ & $M_{3} = 2500\,\GeV$ \\
 &  &  \\
$m_{\bar{f}_{L},\bar{f}_{R}} = 2000\,\GeV$ & $A_{f_{3}} = 1200\,\GeV$ & $\tan\beta = 12$ \qquad $M_{A}=2000\,\GeV$ \\
 &  & \\
$m_{\text{top}} = 173.4\,\GeV$ & $\alpha_{\text{s}}(m_{\text{Z}}) = 0.1181$ & $m_{\tau} = 1.777\,\GeV$ \qquad $m_{\text{b}}(m_{\text{b}})=4.18\,\GeV$ \\
 &  &  \\ \hline
\end{tabular}
\end{table}

\NMT{} uses \code{NMHDECAY}~\cite{Ellwanger:2004xm, Ellwanger:2005dv}
which is based on \code{SDECAY}~\cite{Muhlleitner:2003vg} to compute
the masses, couplings and decay widths of all Higgs bosons and the
masses of all other sparticles. The Higgs spectrum is calculated with
the default settings in the \gnmssm{}, whereas the full two loop
corrections of \(\mathcal{O}(\alpha_s(\alpha_t + \alpha_b))\) are only 
implemented for the \(\mathbb{Z}_3\)-invariant
\nmssm~and the third-generation purely Yukawa corrections are taken 
in the \mssm{} limit. In case of the \nmssm{}
benchmark point presented below, we can compare the numerical
difference in the two setups and find an estimate of the theoretical
uncertainty stemming from missing higher order corrections in the
\gnmssm{} of about \(200\,\MeV\). Thus we conclude that we can safely
use the \NMT{} default configuration for our study. Concerning the
following study, the input values of the couplings $\kappa$,
$\lambda$, and the soft \susy-breaking parameter $A_{\kappa}$, as well
as $\mue$ are varied and \NMT{} calculates the \nmssm{} spectrum for
each point, correspondingly.  We have chosen to scan $\lambda$ and
$\kappa$ between $0$ and $0.1$ each; $|\mue|$ from $100\,\GeV$ to
$1000\,\GeV$; and A$_{\kappa}$ between $-300\,\GeV$ and
$300\,\GeV$. All scanned parameters have been varied uniformly in the
above mentioned intervals where we employed about one million sample
points from which we picked our benchmark scenario. The rather small
range for \(\lambda\) has been chosen explicitly to resemble the
cosmologically relevant parameter region for inflation according to
\cite{Ferrara:2010yw, Ferrara:2010in}, whereas \(|\mue|>100\,\GeV\)
has been chosen to comply with the LEP chargino bound as reported in
\citere{Zyla:2020zbs}. Note that the absence of tachyons in the
spectrum usually requires \(\sign A_\kappa \neq \sign \mue\); we
excluded small absolute values of \(\mue\) to avoid direct exclusion
limits from \lep{} for light charginos.

Concerning the Dark Matter constraints, we have calculated the relic
density and direct detection rates as well as limits from indirect
detection with \NMT{} using \code{micrOMEGAs} version
\code{5.0}~\cite{Belanger:2001fz, Belanger:2005kh, Belanger:2006is,
  Belanger:2013oya, Belanger:2018ccd}. The Dark Matter relic density
is decreased mainly through annihilation of the
\emph{next-to-lightest} supersymmetric particle. An interesting
feature that asks for further investigation. Furthermore, many
observables are calculated and compared with experimental bounds from
LEP and LHC by \NMT{}. Points passing these constraints have then been
checked with \HB{} for \(95\,\%\) C.\,L.\ exclusion at LEP, Tevatron
and LHC; furthermore the \sm-like Higgs properties have been tested
with \HS. We take special emphasis on the Higgsstrahlung process
$e^{+}e^{-}\rightarrow h_{1} \: Z$ which has been important at LEP and
will play the same role at the ILC. For that purpose, we study the
cross section of this process in more detail below and estimate
prospects of a future discovery. The cross section is controlled by
the Higgs coupling to gauge bosons displayed in \eqn{eq:cp}. Finally
we have employed \CM{} to test for current exclusions from Drell--Yan
production at the LHC, as well as neutralino production $p\,p
\rightarrow \tilde{\chi}_{2}^0 \: \tilde{\chi}_{2}^0$, $p\,p
\rightarrow \tilde{\chi}_{1}^0 \: \tilde{\chi}_{2}^0$ and chargino
production $p\,p \rightarrow \tilde{\chi}^{+}_{1} \:
\tilde{\chi}^{-}_{1}$. \CM{} simulates signal events for BSM models at
the LHC and compares with the data from the experimental analyses for
exclusion. As a result, a criterion is provided by \CM{} which is used
to determine whether the parameter point is disfavoured or not. This
criterion is the $r$ value which is defined by the ratio between the
number of simulated signal events $S$ and the 95\% upper limit of
experimental data $S_{95}$:
\begin{equation}
r=\frac{S-1.96\cdot\Delta S}{S_{95}}.
\end{equation}
If $r>1$, the BSM prediction exceeds the 95\% C.\,L.\ and the model is
excluded. Moreover, we calculated cross sections for light Higgs
production $e^{+}e^{-} \rightarrow Z \: h_{1,2}$ using \MG{} version
\code{2.7.2} and display the results below in \figs{fig:xs2}
and~\ref{fig:zh1}. We have identified a benchmark point passing all
experimental constraints implemented in the codes listed above which
comprises a light Higgs at \(97\,\GeV\).

The full mass spectrum of the Higgs, neutralino and chargino sector is
shown in Tab.~\ref{tab:NMSSMspec}. The lightest Higgs has a mass
$m_{h_{1}} = 96.99\,\GeV$, where the \sm{}-like Higgs $m_{h_{2}} =
125.3\,\GeV$. We have accepted \sm{}-like Higgs masses within the
ranges $m_{h_\text{SM}} = (125.1 \pm 3)\,\GeV$ from the scanned points
to select benchmark candidates. Later the \(m_{h_2}\) value is tested
with \HS{} which returns a \(\chi^2\) value of \(86\) with \(107\)
degrees of freedom, including Higgs mass observables, which signals
perfect agreement to a \sm-like Higgs. The heavy \cp-even, \cp-odd and
charged Higgs $H_{3}$, $A$ and $H^{\pm}$ have masses $\lesssim
2000\,\GeV$ as implied by the input value of
Tab.~\ref{tab:NMSSMfixedpar}. The neutralino sector is found to be
slightly above the electroweak scale with the lightest neutralino at
$\sim 190\,\GeV$. However, the second to fourth lightest neutralinos
$\tilde\chi_{2...4}$ are very close in mass to \(\tilde\chi_1^0\)
between $m_{\tilde\chi_{2}} = 194.2\,\GeV$ and $m_{\tilde\chi_{4}} =
255.1\,\GeV$. The nature of the stable Dark Matter candidate is
singlino-like with high purity. It is interesting to notice is that
the next-to-lightest neutralino \({\tilde \chi}_2^0\) is certainly
long-lived to leave any detector in a collider experiment similar to
the Dark Matter. We leave a more detailed study of the Dark Matter
phenomenology of such a scenario for future study. The lightest
chargino has a mass of $m_{\tilde\chi^{\pm}_{1}} = 214.5\,\GeV$ while
the second chargino has the same mass as the heaviest neutralino,
$m_{\tilde\chi^{\pm}_{2}} \approx 2 m_{\tilde\chi^{0}_{5}}$.  The
input parameters of this point as result of the scan are shown in
Tab.~\ref{tab:NMSSMscanpar}. The negative \(\mue\) can be traded for a
negative \(A_\kappa\) without much change. Note, that the large
\(A_\kappa \simeq 270\,\GeV\) is responsible for a heavy \cp-odd
singlet with \(m_a = 273.7\,\GeV\) in contrast to its lighter \cp-even
counterpart.

\begin{table}[tb]
  \caption{Mass spectrum of our \nmssm{} point as given by
    \code{NMSSMTools}. In the Higgs sector we have the lightest scalar
    Higgs $h_{1}$, the \sm-like Higgs $h_{2}$, the Heavy Higgs
    $H_{3}$, as well as the \cp-odd Higgses $a$ and $A$ and the
    charged Higgs $H^{\pm}$. The neutralino sector is labeled with
    $\tilde{\chi}_{1...5}$, and the chargino masses are denoted as
    $m_{\tilde{\chi}^{\pm}_{1,2}}$.}
\label{tab:NMSSMspec}
\centering
\begin{tabular}{lll}
\hline
  &  &  \\
$m_{h_{1}} = 96.99\,\GeV$ & $m_{h_{2}} = 125.3\,\GeV$ & $m_{H_{3}} = 1962\,\GeV$  \\
 &  &  \\
$m_{a} = 273.7\,\GeV$ & $m_{A}= 1962\,\GeV$ & $m_{H^{\pm}} = 1964\,\GeV$   \\
 &  &  \\ \hline
  &  &  \\
$m_{\tilde\chi^0_{1}} = 190.4\,\GeV$ & $m_{\tilde\chi^0_{2}} = 194.2\,\GeV$ & $m_{\tilde\chi^0_{3}} = 226.1\,\GeV$ \\
 &  &  \\
$m_{\tilde\chi^0_{4}} = 255.1\,\GeV$ & $m_{\tilde\chi^0_{5}} = 538.3\,\GeV$ &  \\
 &  &  \\
$m_{\tilde\chi^{\pm}_{1}} = 214.5\,\GeV$ & $m_{\tilde\chi^{\pm}_{2}} = 538.3\,\GeV$ &  \\
 &  &  \\ \hline
\end{tabular}
\end{table}

\begin{table}[tb]
\caption{Results for the parameter scan in the \nmssm{} with $\mue$ at
  the electroweak scale, the soft SUSY-breaking parameter $A_{\kappa}$
  and the couplings $\kappa$ and $\lambda$ leading to the mass
  spectrum shown in Tab.~\ref{tab:NMSSMspec}.}
\label{tab:NMSSMscanpar}
\centering
\begin{tabular}{lll}
\hline
 &  &  \\
 $\mue = -212.3\,\GeV$ & $A_{\kappa} = 268.6\,\GeV$   \\
  &  &  \\
 $\kappa = 0.01846$ & $\lambda = 0.04215$   \\
  &  &  \\ \hline
\end{tabular}
\end{table}

\subsection{\munmssm{} study of the effects from \(\mui\)}
\label{sec:muinfeff}
Starting from the benchmark point discussed above, we are interested
to see the effect of $\mui$. The \nmssm{} limit is reached for \(\mui
= 0\,\GeV\). We increase the value of \(\mui\) from \(0\) to
\(1000\,\GeV\) and study how the spectrum is changed, how the mixing
is affected, and finally how the phenomenology (reduced couplings and
branching ratios) of the light Higgs states vary under modulation of
\(\mui\). All the other parameters are kept the same.

\begin{figure}[tp]
  \centering
  \includegraphics[width=.9\textwidth]{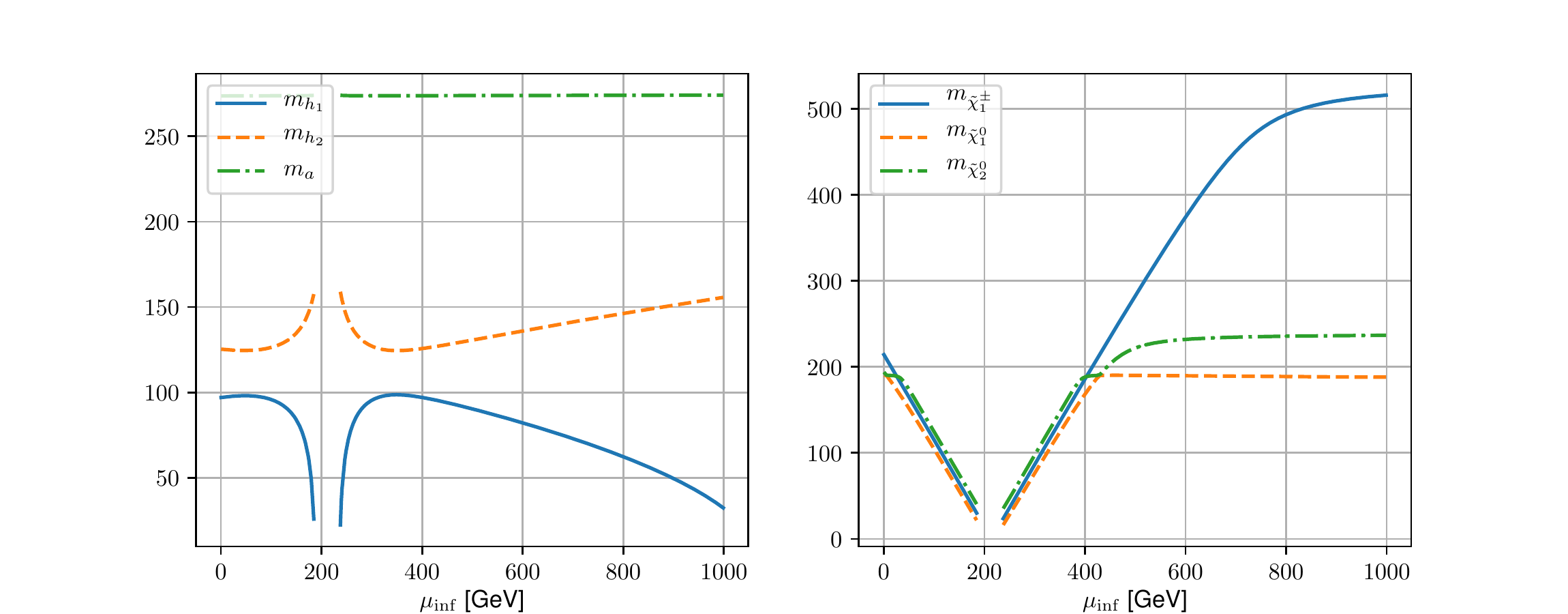} 

  \caption{\label{fig:ms1} The masses of scalar higgses $h_1$, $h_2$,
    pseudoscalar higgs $a$, neutralinos ${\tilde\chi}^0_1$,
    ${\tilde\chi}^0_2$, and chargino ${\tilde\chi}^\pm_1$, depending
    on $\mui$.}

  \vspace{2ex} \hrule \vspace{2ex}

  \capstart

  \includegraphics[width=.8\textwidth]{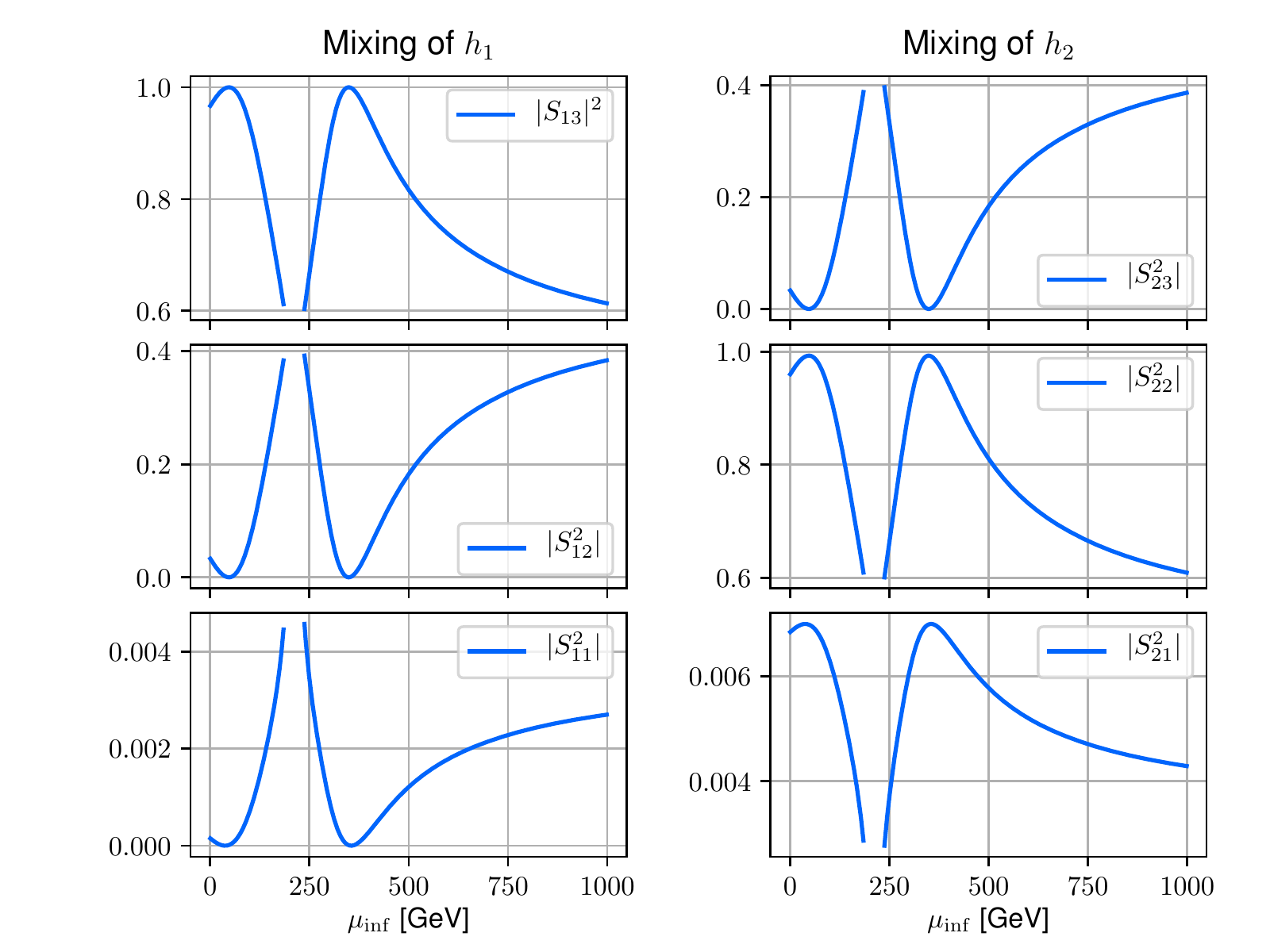} 
  \caption{\label{fig:mx1} The three mixing components of two lightest
    Higgs bosons depending on $\mui$. From bottom to top: the down
    type components are $|S_{11}^2|$ and $|S_{21}|^2$, the up type
    components are $|S_{12}|^2$ and $|S_{22}|^2$, and the singlet
    components are $|S_{13}|^2$ and $|S_{23}|^2$.}

  \vspace{2ex} \hrule

\end{figure}

We show the spectrum of the light \cp-even Higgs bosons \(h_{1,2}\)
and the light \cp-odd state \(a\), as well as the light neutralinos
and charginos in~\fig{fig:ms1}. For \(\mui = 0\,\GeV\) we recover the
\nmssm{} spectrum given in Tab.~\ref{tab:NMSSMspec}. Around \(\mui =
200\,\GeV\) the mass of the light Higgs \(h_1\) turns into a tachyonic
dip where no line is shown and finally rises again towards \(\mui
\simeq 348\,\GeV\), where it reaches a second maximum. This is an
amusing feature observed in the numerics and shows how in this model
parameter points with a similar mass spectrum although having
distinguished fundamental parameters can be achieved. At around \(\mui
\approx 210\,\GeV\), the combination \(\mue + \mui\) is close to zero,
which drives the tachyonic behaviour. The first maximum, corresponding
to the first minimum of \(m_{h_2}\) is around \(\mui=46\,\GeV\). In
contrast to this rich evolution of \(m_{h_{1,2}}\) with \(\mui\), the
mass of \(a\) varies only mildly and is dominated by the fixed value
of \(A_\kappa\). On the right hand side of \fig{fig:ms1}, we show the
light neutralino masses evolving with \(\mui\). In the regime below
\(400\,\GeV\), all three displayed masses behave linearly with
\(\mui\), where for larger \(\mui \gtrsim 400\,\GeV\) the dominant
wino-, bino-, and singlino-like behaviour is developed. The linearly
rising mass with \(\mui\) belongs to higgsino-like states, as their
mass is mainly driven by \(\mue +\mui\). The singlino, in contrast is
supposed to stay constant under variation of \(\mui\) as can be seen
from the mass matrix in \eqn{eq:neutralinomassmatrix}, where
\(\left(M_{{\tilde\chi}^0}\right)_{55} = 2 \frac{\kappa}{\lambda} \mue
= - 185.958\,\GeV\) for the parameters in this scenario given in
Tab.~\ref{tab:NMSSMscanpar}. This shows how differently the spectra of
Higgs bosons and neutralinos/charginos evolve with \(\mui\). Although
there are three distinct values of \(\mui\) where the Higgs spectrum
essentially looks the same as for the \nmssm{} point, for two of them
the neutralinos become much lighter and thus in conflict with Dark
Matter phenomenology. We have identified one point at \(\mui \simeq
395\,\GeV\) which comprises the same spectra for both Higgs and
neutralino/chargino as for \(\mui = 0\,\GeV\).

\begin{figure}[tp]
  \centering

  \includegraphics[width=\textwidth]{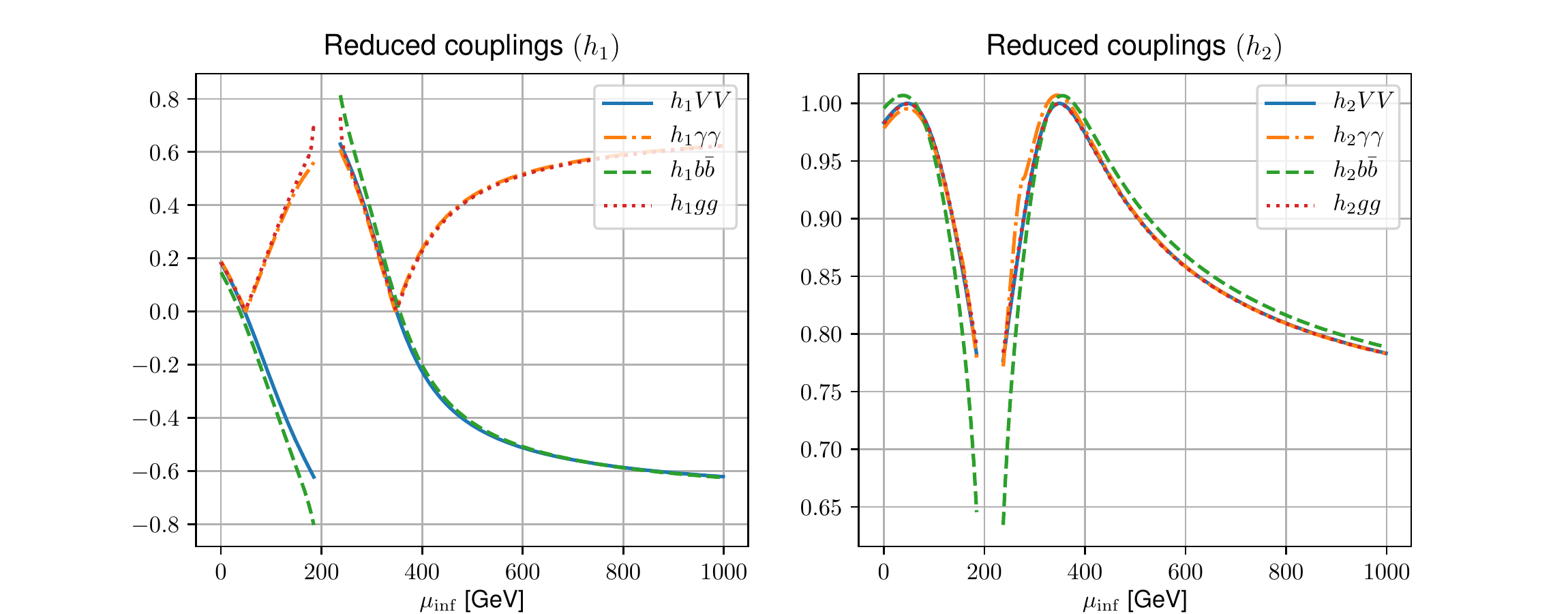} 

  \caption{\label{fig:rc1} The reduced couplings of $h_1$ and $h_2$ to
    gauge bosons, photons, $b$ quarks and gluons, depending on $\mui$.}

  \vspace{2ex} \hrule \vspace{2ex}

  \capstart

  \includegraphics[width=.8\textwidth]{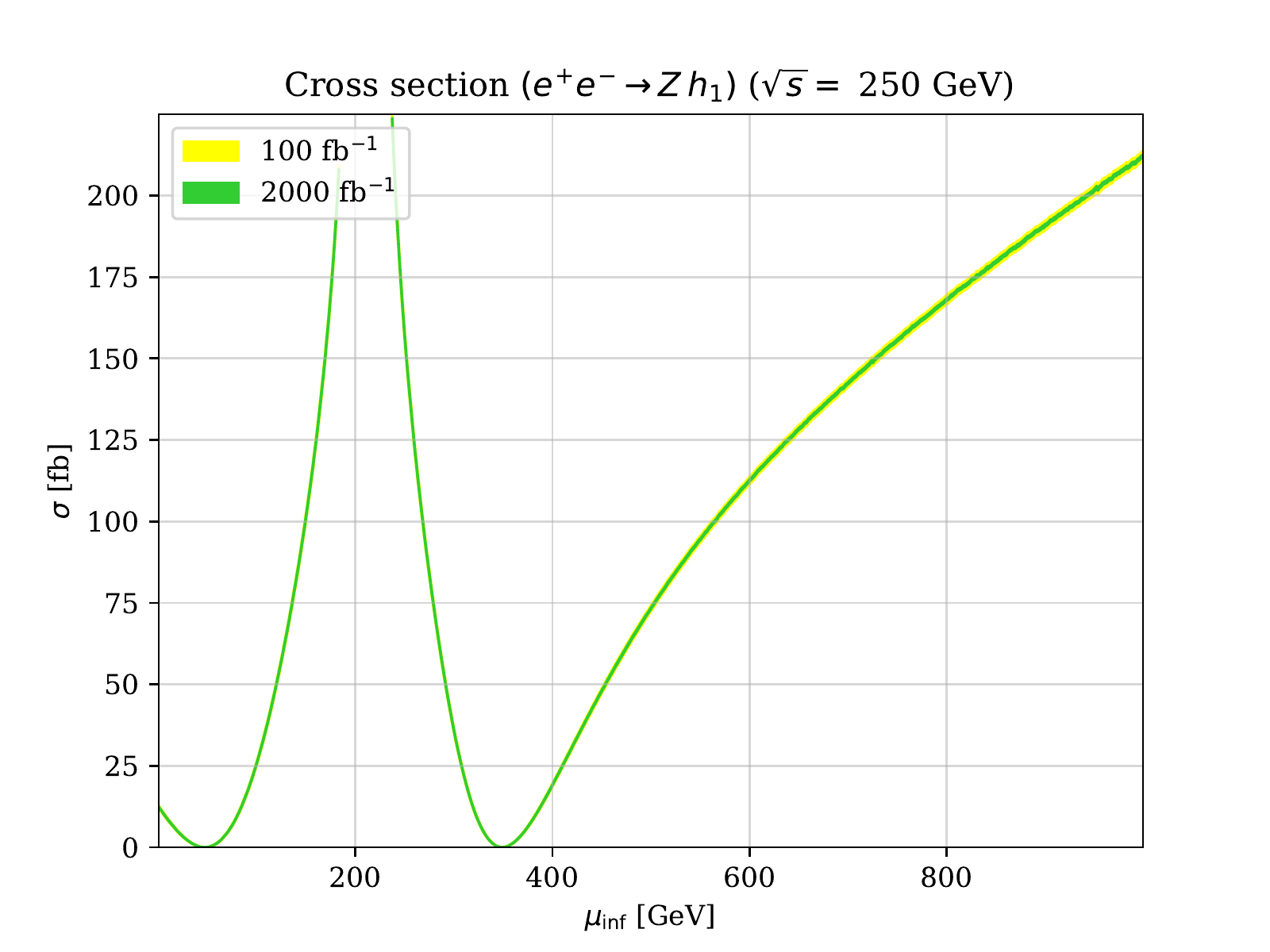} 

  \caption{\label{fig:xs2} The lightest scalar Higgs production cross
    section at 250 GeV ILC depending on $\mui$.}

  \vspace{2ex}
  \hrule
\end{figure}

Crucial for the phenomenology of this scenario is a view on the Higgs
mixing matrices, especially the singlet-doublet mixings as shown in
\fig{fig:mx1}. Here, we show the singlet admixture to the lightest
state (left side top), and the doublet components of the same (left
side middle and down). On the right hand side, the same is shown for
the second lightest state. It is interesting to see that there are two
degenerate points, where \(h_1\) is purely singlet and \(h_2\) purely
doublet. These points coincide with the minima and maxima in the
spectrum of \fig{fig:ms1}. Towards large values of \(\mui\), the
second lightest Higgs becomes singlet-dominated, while the lightest
loses its singlet character. Note, however, that there is no scalar at
\(125\,\GeV\) in the spectrum anymore, so the regime of large \(\mui\)
is disfavoured by observations.

The Higgs mixing also defines the reduced couplings at the tree level,
see \eqns{eq:topbotcoup} and \eqref{eq:cp}. The reduced couplings as
delivered by \code{NMSSMTools} are shown in \fig{fig:rc1}, where we
display the reduced couplings to electroweak gauge bosons (\(VV\)),
photons (\(\gamma\gamma\)), bottom quarks (\(b\bar b\)), and gluons
(\(gg\)) for the lightest and second lightest Higgs, \(h_1\) and
\(h_2\) respectively. It can be seen that for the two points mentioned
above with \(\mui \simeq 46\,\GeV\) and \(\simeq 348\,\GeV\) the
reduced couplings of \(h_2\) approach the \sm{} values, where in
contrast the couplings of \(h_1\) turn to zero. This is exactly the
pure singlet case. In the neighbouring regime, the singlet-like state
has small couplings to the \sm{} and the couplings of \(h_2\) deviate
from the \sm{} values. It is furthermore interesting to notice that
the reduced couplings of the lightest state \(h_1\) to gauge bosons
and bottom quarks have the same absolute value but opposite signs in
the regime \(46\,\GeV \lesssim \mui \lesssim 348\,\GeV\). This gives a
handle to distinguish finally the two degenerate spectra for different
values of \(\mui\). Especially for the point degenerate with the
\nmssm{} case as discussed above for \(\mui = 395\,\GeV\), the reduced
couplings to \(b\) quarks and vector bosons have the opposite sign
while the whole spectrum is identical. This reduced couplings can be,
to some extend, identified with the coupling modifiers in the
\(\kappa\) framework for \sm{} Higgs studies, as pointed out in
Sec.~\ref{sec:Higgs}. This becomes more relevant in the following
section, where we study a scenario with a very \sm-like Higgs over the
full \(\mui\) range.

The couplings to gauge bosons, especially the \(Z\) boson, also define
the behaviour of the production cross section at a lepton collider,
such as the ILC, in the dominant production mode via
Higgsstrahlung. We display in~\fig{fig:xs2} how the cross section for
\(e^+e^- \to Z\, h_1\) evolves with \(\mui\) in this scenario for an
initial center of mass energy \(\sqrt{s} = 250\,\GeV\). Of course, the
pure singlet case at \(\mui = 48\,\GeV\) and \(348\,\GeV\) cannot be
produced. With a certain doublet admixture, however, a light
singlet-like state can be produced at the ILC250 with a few femtobarn
cross section. The coloured bands show the statistical uncertainties
for integrated luminosities of \(L=100/\mathrm{fb}\) (yellow) and
\(L=2000/\mathrm{fb}\) (green). The cross section uncertainty is
derived as statistical uncertainty from a counting analysis:
\begin{equation}
\delta\sigma=\frac{\sigma}{\sqrt{N}}=\sqrt{\frac{\sigma}{L}},
\end{equation}
where the Poisson distribution defines the uncertainty from the number
of signal events as \(\sqrt{N}\). A delicate analysis of the discovery
potential is beyond the scope of this paper. The simplified procedure
described above for an estimate relies on a theoretical prediction
under the assumption of a perfect experiment and thus neglecting
detector effects.

\begin{figure}[tb]
  \centering
  \includegraphics[width=\textwidth]{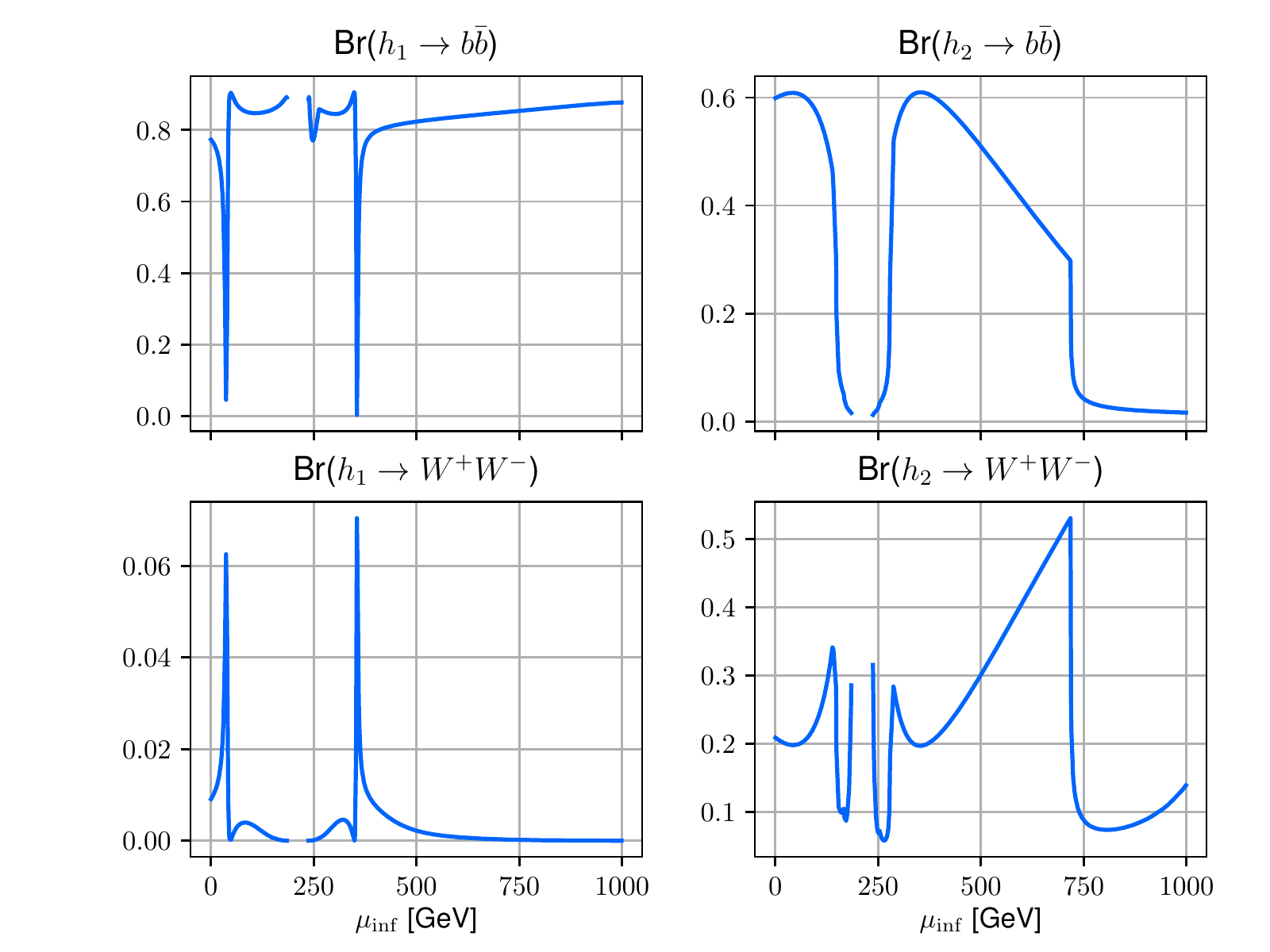}
  \caption{The branching ratios of $h_1$ and $h_2$ decay to $b$ quarks
    or $W$ bosons depending on $\mu_{\text{inf}}$.}
  \vspace{2ex} \hrule
  \label{fig:br1}
\end{figure}

Finally, we show the branching ratios for decays to bottom quarks and
\(W\) boson pairs in~\fig{fig:br1}. The light state \(h_1\) mainly
decays to bottom quarks over most of the displayed \(\mui\)
range. Only at the points where it becomes exclusively singlet, the
branching ratio to bottom quarks drops towards zero. For the second
lightest state, the branching ratio to bottom quarks also goes down in
the interval \(125\,\GeV \lesssim \mui \lesssim 275\,\GeV\), which is
partially compensated by an increase in decays to \(W\) bosons. For a
more detailed study of the behaviour, all decay modes have to be
included. The rapid decrease of branching fractions of \(h_2\) into
both \(b\bar b\) and \(W\) pairs at below \(\mui\simeq 750\,\GeV\) is
due to the opening of the \(h_2 \to h_1 h_1\) decay channel, where
\(m_{h_2}\) becomes twice \(m_{h_1}\). The displayed branching ratios
of \(h_2\) go down in the window around \(\mui \simeq 200\,\GeV\)
because here the decays into neutralinos and charginos become relevant
(notice their corresponding small masses in this window). The two dips
in \(\operatorname{Br}(h_2 \to W^+ W^-)\) are due to an enhanced
\(\operatorname{Br}(h_2 \to {\tilde\chi}_1^0 {\tilde\chi}_2^0)\) in
these regimes.

\subsection{Reweighting $\mui$ effects in the spectrum} \label{sec:muistudy}
It has been remarked in a previous study of the inflationary \munmssm,
\citere{Hollik2019}, that the neutralino spectrum at the tree level
stays invariant under changes of \(\mui\) when the singlet
self-coupling \(\kappa\) is adjusted appropriately. Under the same
redefinition also the scalar spectrum does not change over vast
regions in the parameter range aside from extreme configurations. Such
an extreme case has been discussed in \citere{Hollik2019}. In the
following, we refrain from artificial cancellations in the mass
matrices and choose rather combinations of parameters to be constant
such that variations in \(\mui\) enter mildly. From a quick study of
the scalar mass matrix given in~\eqns{eq:mx1}, we see that three
combinations are dominantly controlling the matrix elements. One is
the sum \(\mue + \mui\), then we have \(\frac{\kappa}{\lambda} \mue\)
repeatedly appearing and furthermore the combination that has been
replaced by the charged Higgs mass dominating the heavy doublet mass
eigenvalue.

We treat the following combinations constant under variation of
\(\mui\), which implies a redefinition of \(\kappa\) and \(\mue\):
\begin{subequations}
\begin{align}
a=&\; \mui + \mue\,, \\
b=&\; \frac{\kappa}{\lambda} \mue\,, \\
c=&\; \mue (\frac{\kappa}{\lambda} \mue + A_\lambda)
\equiv \frac{1}{2} (m_{H^\pm}^2-m_W^2+v^2\lambda^2) \sin 2\beta\,.
\end{align}
\end{subequations}

Keeping these combinations fixed, under variation of \(\mui\) the
upper left blocks of the Higgs mass matrices are unchanged. The other
mass matrix elements with a residual \(\mui\) dependence can then be
expressed as
\begin{subequations} \label{eq:newSmass}
\begin{align}
M_{S,33}^2=&\; \lambda^2 v^2 \left( \frac{\cos\beta\sin\beta}{a-\mui}
(\frac{c}{a-\mui} - b) - \frac{\mui}{a-\mui} \right) +
b(A_\kappa+4b)\,, \\
M_{S,13}^2=&\; M_{S,31}^2 =\; v\lambda \left(
2a\cos\beta-(\frac{c}{a-\mui}+b)\sin\beta\right)\,,
\\ M_{S,23}^2=&\;
M_{S,32}^2 =\; v\lambda \left(
2a\sin\beta-(\frac{c}{a-\mui}+b)\cos\beta\right)\;,
\end{align}
\end{subequations}
and
\begin{subequations} \label{eq:newPmass}
\begin{align}
M_{P,33}^2=&\; \lambda^2 v^2 \left( \frac{\cos\beta\sin\beta}{a-\mui}
(3b+\frac{c}{a-\mui}) - \frac{\mui}{a-\mui} \right)\;, \\
M_{P,13}^2=&\; M_{P,31}^2 =\; -v\lambda \left(
3b-\frac{c}{a-\mui}\right) \sin\beta\;, \\
M_{P,23}^2=&\; M_{P,32}^2=\;
-v\lambda \left( 3b-\frac{c}{a-\mui}\right)\cos\beta\;.
\end{align}
\end{subequations}

\begin{figure}[tb]
        \centering
        \includegraphics[width=\textwidth]{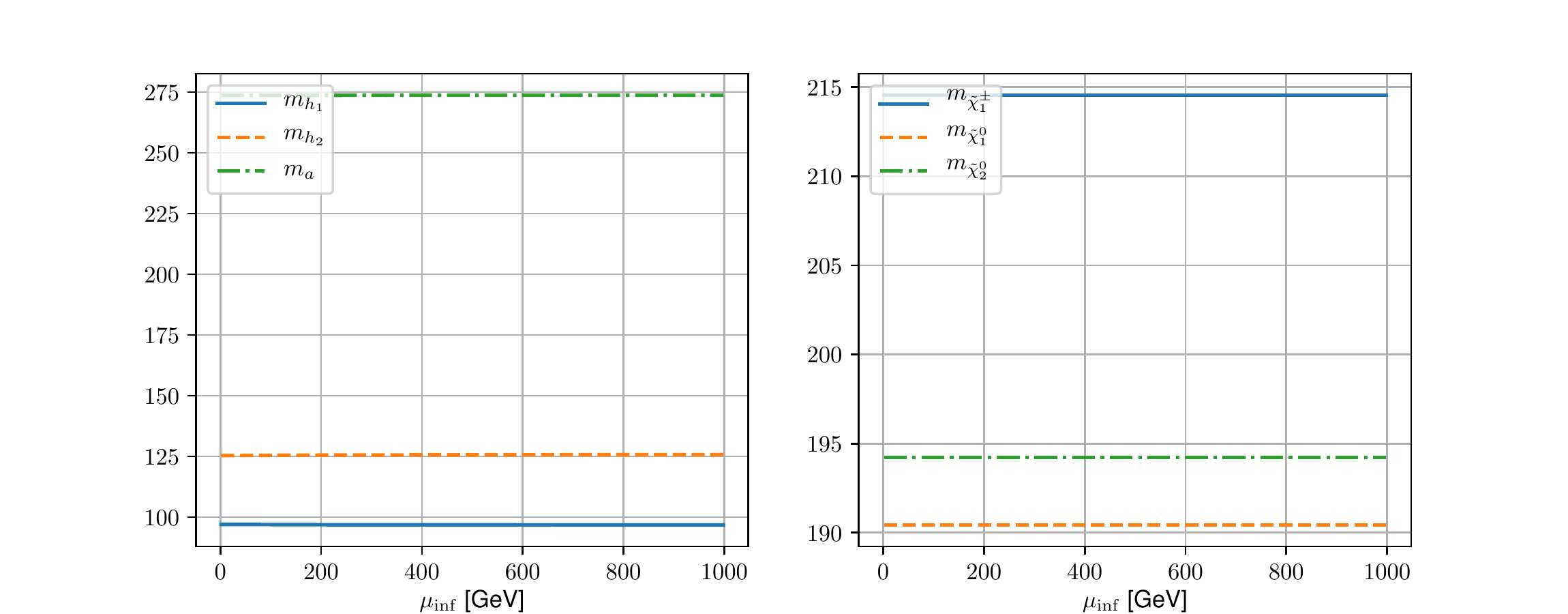}
        \caption{The masses of the light \cp-even states $h_1$, $h_2$,
          and the \cp-odd singlet-like state $a$ (left); the masses of
          the light neutralinos ${\tilde\chi}^0_1$,
          ${\tilde\chi}^0_2$, and the light chargino
          ${\tilde\chi}^\pm_1$ (right), depending on the pure $\mui$
          effect.}
        \vspace{2ex} \label{fig:ms2}
\hrule
\end{figure}

Note, that the parameters $\lambda$, $A_\kappa$, and $\tan\beta$ can
be essentially varied without changing the fixed combinations from
above. Since we are studying the pure effect of \(\mui\) while
minimally invasively changing the mass spectrum, we also keep them at
the values specified in Tab.~\ref{tab:NMSSMscanpar}, where \(\kappa\)
is not kept at that value. This can be seen also from \eqns{eq:newSmass}
and~\eqref{eq:newPmass} where the appearance of \(\kappa\) is
absorbed. The mass spectrum is then only slightly changing under
increase of \(\mui\) from \(0\,\GeV\) to \(1000\,\GeV\) in contrast to
what has been shown in Sec.~\ref{sec:muinfeff}. We show the
correspondance of \fig{fig:ms1} in \fig{fig:ms2}.

\begin{figure}[tp]

  \centering

  \includegraphics[width=\textwidth]{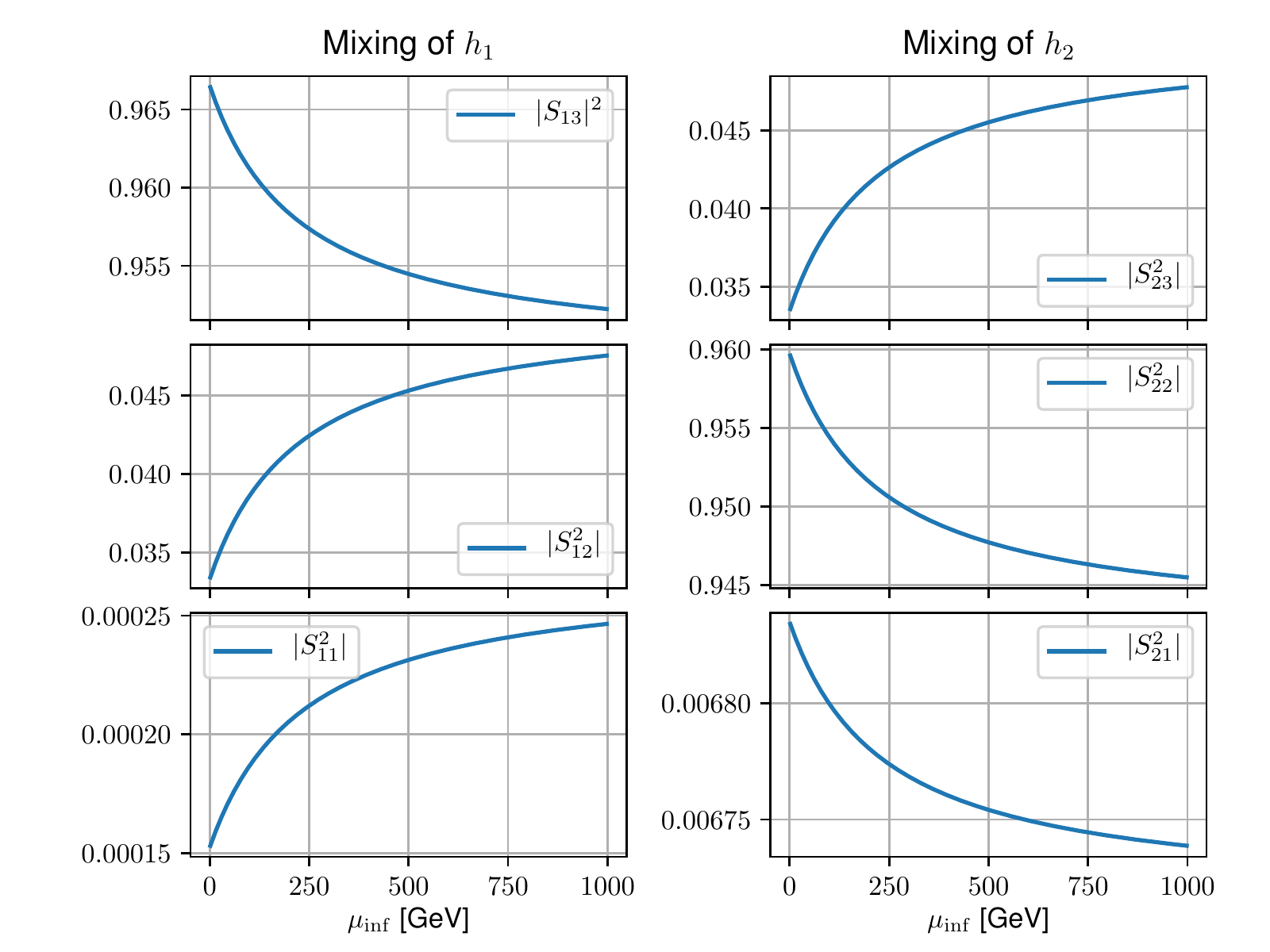}

  \caption{The three mixing components of the lightest scalar Higgses
    $h_{1,2}$ depending on the pure $\mui$ effect. The down type component
    of $h_1$ is $|S_{11}^2|$, the up type component of $h_1$ is
    $|S_{12}^2|$ and the singlet component of $h_1$ is
    $|S_{13}^2|$.} \label{fig:mix}

  \vspace{2ex} \hrule \vspace{2ex}

  \capstart

  \includegraphics[width=\textwidth]{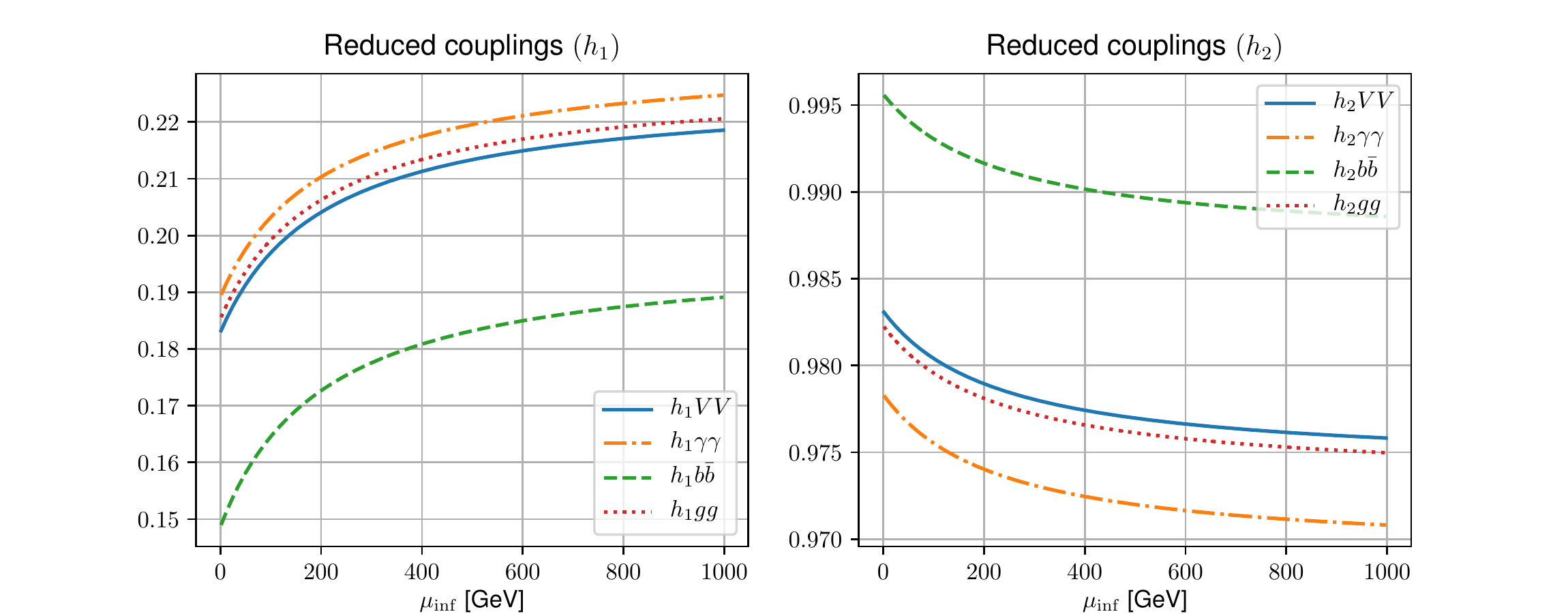}

  \caption{The reduced couplings of $h_1$ and $h_2$ to gauge bosons,
    photons, $b$ quarks and gluons, depending on the pure
    $\mu_{\text{inf}}$ effect.} \label{fig:rc2}

  \vspace{2ex} \hrule
\end{figure}

The question is now, how much the phenomenology of a \munmssm{} point
with large \(\mui\) differs from a point close to the \nmssm{}
limit. Taking a look at the Higgs mixing components in \fig{fig:mix},
we see that the singlet admixture to the lightest state only mildly
decreases. All changes in the mixings are less than at most
\(15\,\%\).  It is interesting to notice that for increasing \(\mui\),
the doublet admixture to the lightest Higgs increases, where
simultaneously the doublet components in \(h_2\) become less
relevant. Moreover, the larger \(\mui\) the less rapid the change.

The behaviour of the mixing components with respect to \(\mui\) is
also mirrored in the reduced couplings shown in
\fig{fig:rc2}. Measuring a deviation of less than \(2\,\%\) from the
\sm-values for the \sm-like scalar is more than challenging at the
\lhc{} and any future collider. Increasing \(\mui\) to around
\(1\,\TeV\), we would have a deviation of less than \(3\,\%\) for the
coupling to photons, where the bottom quark coupling of \(h_2\)
deviates only a bit more than \(1\,\%\) from the \sm. Since for larger
\(\mui\) the curves flatten out, a further increase of \(\mui\) in
this scenario does not give a sizeable effect. On the other hand, the
singlet-like scalar \(h_1\) shows couplings of around \(15-20\,\%\) of
a \sm-Higgs at the same mass of \(97\,\GeV\). That means, if
non-vanishing couplings can be measured to more than \(10\,\%\) at a
future collider, there is a clear discovery potential for this
singlet-like state. Nevertheless, it looks less promising to
distinguish the \munmssm-scenario from the \nmssm{} point by just
comparing the reduced couplings in the \(\kappa\) framework. If we
look \EG{} on the \(h_2\) coupling to vector bosons in \fig{fig:rc2}
(the blue continuous curve), which can be identified with
\(\kappa_V\), there is a variation of less than \(0.01\) over the
displayed range. Supposed that at the ILC this \(\kappa_V\) can be
measured to more than \(1\,\%\) accuracy~\cite{Waescheleine}, a
deviation might be visible. The corresponding measurements of the
signal strength for the singlet-like state, however, look more
promising.

\begin{figure}[tp]

  \centering

  \includegraphics[width=\textwidth]{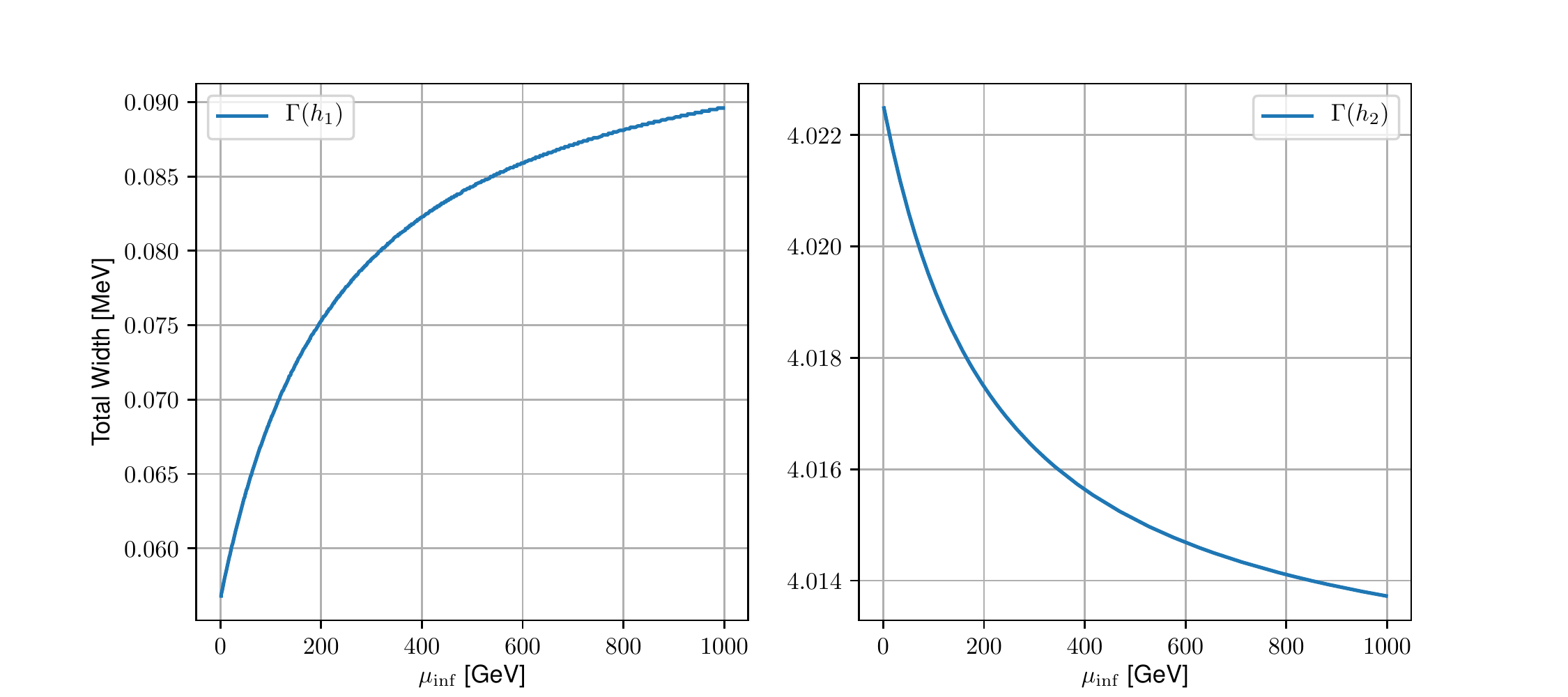}

  \caption{The total widths of the lightest scalar Higgs $h_1$ and
    second lightest Higgs $h_2$ depending on the pure
    $\mu_{\text{inf}}$ effect.} \label{fig:width}

  \vspace{2ex} \hrule \vspace{2ex}

  \capstart

  \includegraphics[width=.8\textwidth]{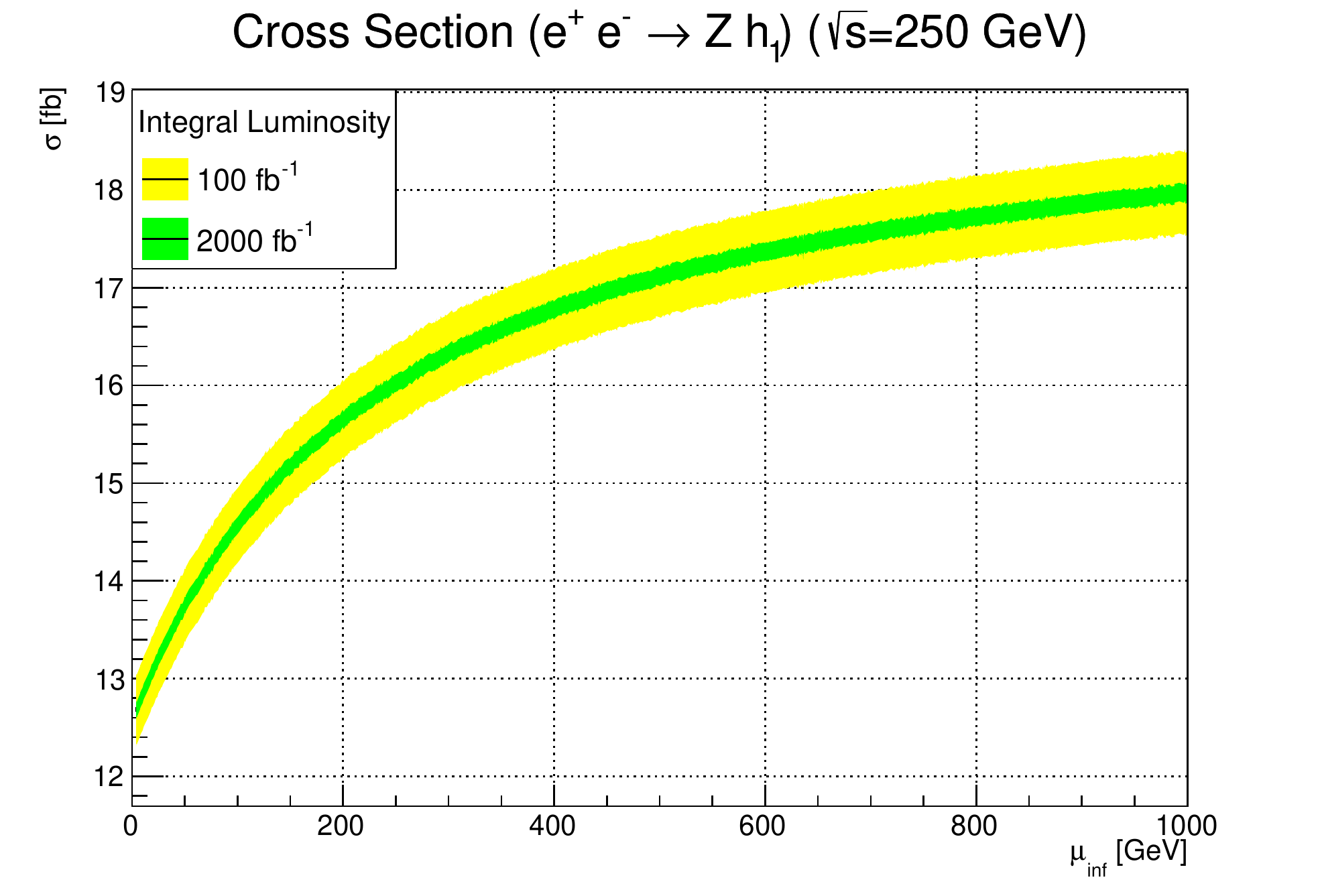}

  \caption{The lightest scalar Higgs production cross section at 250
    GeV ILC depending on the pure $\mu_{\text{inf}}$ effect.}
  \label{fig:zh1}

  \vspace{2ex} \hrule
\end{figure}

The same effect can also be seen in the total widths of \(h_1\) and
\(h_2\) displayed in \fig{fig:width}, where the curves follow the
behaviour of the reduced couplings. Due to the rather small total
width of the lightest Higgs boson, the effect of an increasing
\(\mui\) is very prominent here, where the total width is nearly
doubled over the displayed range. In contrast, for \(h_2\) the total
width is only mildly affected and its variation probably out of
reach. Since we are on top of the \sm-value for the total width around
\(4\,\MeV\), see \citeres{deFlorian:2016spz, Heinemeyer:2013tqa},
there is also not much room for invisible decay modes that are also
not predicted in this scenario.

For a future study of this model at a collider, especially an \(e^+
e^-\) machine, the production cross section of the singlet-dominated
state is important. We calculate the cross section in Higgstrahlung at
the ILC for a center of mass energy \(\sqrt{s} = 250\,\GeV\) as in
Sec.~\ref{sec:muinfeff}. The result over the range \(\mui \in [0,
  1000]\,\GeV\) is shown in \fig{fig:zh1}. Starting from the \nmssm{}
benchmark point with \(\mui = 0\,\GeV\) and a cross section of about
\(12.6\,\mathrm{fb}\), the total cross section is enhanced by about
\(50\,\%\) at \(\mui = 1000\,\GeV\). Already for \(\mui = 200\,\GeV\)
there is an increase of one quarter with respect to the initial cross
section in the pure \nmssm{} scenario. In general, we want to stress
that cross sections of more than \(10\,\mathrm{fb}\) are well in reach
for a linear collider
\cite{Fujii:2017ekh,Moortgat-Picka:2015yla,Baer:2013cma}. A cross
section enhanced by \(50\,\%\) compared to the \nmssm{} case is a
clear sign of a possible distinction. The yellow and green coloured
bands in \fig{fig:zh1} show the statistical uncertainties after an
integrated luminosity of \(100/\mathrm{fb}\) and \(2000/\mathrm{fb}\),
respectively. The interpretation of these uncertainty bands is most
useful when distinguishing two parameter points for different values
\(\mui\). At \EG{} \(\mui = 200\,\GeV\) the uncertainty band allows
for cross sections between \(15.2\) and \(16\,\mathrm{fb}\) with
\(100/\mathrm{fb}\) of recorded data. Similarly, a cross section of
\(16\,\mathrm{fb}\) hints of a \(\mui\) in the range between \(200\)
and \(300\,\GeV\). Nevertheless, for small values of \(\mui\), the
uncertainties are also smaller in absolute terms and a \(\mui\) of
\(50\,\GeV\) can be clearly distinguished from the \(\mui = 0\,\GeV\)
case.  If we assume that the ILC can reach an integral luminosity of
up to \(2000/\mathrm{fb}\), the statistical uncertainty is narrowed
down giving a much higher potential for distinction.  In this case a
measured cross section can be assigned to a smaller range of $\mui$
and conversely larger values of $\mui$ could be distinguished at the
experiment. Note that we have considered the statistical error only
for the displayed cross section, especially we did not consider the
detection efficiency and possible backgrounds in the experimental
study. However, we believe that the ILC at \(250\,\GeV\) has a clear
potential to distinguish the \munmssm{} from the \nmssm{} as well as
certain scenarios within the same model and encourage further
experimental studies including detector effects.

\section{Conclusions} \label{sec:concl}
We have studied in detail the electroweak phenomenology of a
supersymmetric model which incorporates inflation in the early
universe. The model has the same particle content as the \nmssm{} and
comprises an additional singlet superfield. In contrast to the \nmssm,
the speciality of our model is an additional \(\mu\)-term like in the
\mssm{} originating from the non-minimal coupling to gravity, leading
to the so called \munmssm. Our study is focused on properties of the
Higgs sector with a special emphasis on a light singlet-like state at
\(97\,\GeV\). We have presented two routes how to distinguish a
parameter point in the \munmssm{}---where \(\mui\) is the parameter
relevant for inflation---from a corresponding parameter point in the
\nmssm. The benchmark point in the \nmssm{} has been chosen from a
random scan over \nmssm-specific parameters obeying all current
experimental constraints.

For the numerical study, we have employed the public code collection
\NMT{} which serves as spectrum generator and calculates several
observables. \NMT{} does not provide the input options for the
\munmssm, so we had to redefine the parameters in an appropriate way
adopting the code for our model. We have identified a benchmark
scenario to study the phenomenological differences of the \nmssm{} and
the \munmssm{}. This benchmark scenario provides an allowed parameter
point in the \nmssm, where we have checked against existing collider
physics constraints by the use of \HB/\HS{} and \CM. Starting from
this valid point with \(\mui = 0\,\GeV\), we have studied the full
effect of \(\mui \neq 0\,\GeV\) to see how the spectrum and the mixing
changes once this parameter is turned on. We have found a drastic
influence on the mass spectrum, especially with one region where the
lightest Higgs states turns to be tachyonic. Over the full range of
\(\mui\) we have identified one more parameter point where the mass
spectrum of Higgs bosons and neutralinos/charginos is degenerate with
the \nmssm{} point. However, taking a look at the reduced couplings of
the singlet-like state to electroweak gauge bosons and bottom quarks,
we see a difference in the sign which may give a potential for
discinction of the two models. Furthermore, we have calculated the
production cross section of the lightest Higgs in Higgsstrahlung at
the ILC with a center of mass energy \(\sqrt{s} = 250\,\GeV\). For the
relevant physical points it is around \(10\,\mathrm{fb}\) and offers
the possibility for a detailed study at the linear collider.

As a second route to study the ``pure'' \(\mui\) effect, we have
reweighted other parameters to keep the mass spectrum invariant under
variations of \(\mui\). Even in this case, there is a sizeable effect
on the Higgs mixing of a few percent and a reduction of the reduced
couplings of the \sm-like Higgs state to \sm{} particles. Although the
reduced couplings (or coupling-strength modifiers \(\kappa\)) deviate
only by a few percent from the \sm-value, such small deviations will
be measureable at the future linear collider. In contrast, the
singlet-like \cp-even state at \(97\,\GeV\) receives enhanced
contributions to the couplings to \sm-particles due to an enhanced
doublet admixture. Here, the change for increased \(\mui\) is more
prominent with several percent. It is important to notice that the
reduced couplings of the singlet-like state with respect to a
\sm-Higgs at \(97\,\GeV\) are about \(20\,\%\) and therewith
sufficiently large. The Higgsstrahlung cross section of the lightest
Higgs at ILC250 is also increasing with increasing \(\mui\) reaching
\(18\,\mathrm{fb}\) in the scenario under scrutiny. This offers the
possibility to distinguish the \nmssm{} and \munmssm{} scenarios from
a measurement of the production cross section with sufficient integral
luminosity.

\small{
\section*{Acknowledgments}
We would like to thank Cyril Hugonie and Ulrich Ellwanger for their
very helpful communication about the use of \NMT. C.\,L., G.\,M.\,P,
and S.\,P.\ acknowledge the support by the Deutsche
Forschungsgemeinschaft (DFG German Research Association) under
Germany's Excellence Strategy --EXC 2121 ``Quantum Universe''--
390833306. W.\,G.\,H.\ is partially supported by the collaborative
research center TRR 257 ``Particle Physics Phenomenology after the
Higgs Discovery''. We thank Sven Heinemeyer, Georg Weiglein, Stefan
Liebler, Sebastian Pa\ss{}ehr for valuable discussions; furthermore, 
we thank Sven Heinemeyer and Sebastian Pa\ss{}ehr for a thorough read 
and comments on the manuscript.}

\bibliographystyle{utcaps}
\bibliography{reference}
\end{document}